%% file: main.tex
\documentclass[twocolumn]{openjournal}
\usepackage{graphicx} % Required for inserting images

\usepackage[colorlinks,linkcolor=blue,citecolor=blue,urlcolor=blue ]{hyperref}
\usepackage[utf8]{inputenc}
\usepackage{float}
\usepackage{xcolor}
\usepackage{ulem}
\usepackage[T1]{fontenc}

\makeatletter

\newcommand{\Rmnum}[1]{\expandafter\@slowromancap\romannumeral #1@}

\newcommand{\gsim}{\lower0.6ex\vbox{\hbox{$\buildrel{\textstyle >}\over{\sim}\ $}}}

\def\hmsun{{h^{-1} M_{\odot}}}
\def\msun{\, M_{\odot}}
\def\astrid{\texttt{ASTRID} }

\begin{document}
\title[The first few orbits are key to MBH sinking]
{MAGICS I. The First Few Orbits Encode the Fate of Seed Massive Black Hole Pairs}

\author{Nianyi Chen,$^{1,*}$}
\author{Diptajyoti Mukherjee,$^{1}$}
\author{Tiziana Di Matteo,$^{1,2}$}
\author{Yueying Ni,$^{3}$}
\author{Simeon Bird,$^{4}$}
\author{Rupert Croft$^{1,2}$}
\thanks{$^*$E-mail:nianyic@andrew.cmu.edu}

\affiliation{
$^{1}$ McWilliams Center for Cosmology, Department of Physics, Carnegie Mellon University, Pittsburgh, PA 15213 \\
$^{2}$ NSF AI Planning Institute for Physics of the Future, 
Carnegie Mellon  University, Pittsburgh, PA 15213, USA \\
${3}$ Harvard-Smithsonian Center for Astrophysics, 60 Garden Street, Cambridge, MA 02138, USA \\
${4}$ Department of Physics and Astronomy, University of California Riverside, 900 University Ave, Riverside, CA 92521}

\begin{abstract}
The elusive massive black hole (MBH) seeds stand to be revealed by the Laser Space Antenna Interferometer through mergers. 
As an aftermath of galaxy mergers, MBH coalescence is a vastly multi-scale process connected to galaxy formation.
We introduce the ``Massive black hole Assembly in Galaxies Informed by Cosmological Simulations" (MAGICS) suite, with galaxy/MBH properties and orbits recovered from large-volume cosmological simulation ASTRID.
The simulations include subgrid star formation, supernovae feedback, and MBH accretion/feedback. 
In this first suite, we extract fifteen representative galaxy mergers with seed MBHs to examine their dynamics at an improved mass and spatial resolution (by $\sim2000$ and $\sim20$) and follow MBH orbits down to $\sim10\,\text{pc}$.
We find that the seed MBH energy loss and orbital decay are largely governed by global torques induced by the galaxy merger process on scales resolvable by cosmological simulations.
Specifically, pairs sink quickly if their orbits shrink rapidly below $1\,\text{kpc}$ during the first $\sim200\,\text{Myr}$ of pairing due to effective energy loss in major galaxy mergers, whereas MBHs gaining energy in minor galaxy mergers with close passages are likely to stall.
High initial eccentricities ($e_{\rm init}>0.5$) and high stellar densities at kpc scales ($\rho_{\rm star}>0.05\,M_\odot/\text{pc}^3$) also lead to most efficient decays.
$\sim50\%$ high-redshift seed MBH pairs experience consecutive galaxy mergers and are more likely to stall at $\sim1\,\text{kpc}$.
For a subset of systems, we carry out N-Body re-simulations until binary formation and find that some systems merge at high-z when embedded in sufficient nuclear star clusters.\\[0.5em]
%\tiziana{can you fit a short sentence on the mean timescale for seed mergers --> redshift we expect them to 'detect' them with LISA}\\[1em]
\end{abstract}

\maketitle

\input{Sec1_Intro.tex}
\input{Sec2_Method.tex}

\input{Sec3_Population}

\input{Sec4_ResimIC}

\input{Sec5_Result}

\input{Sec6_Conclusion}

\section*{Acknowledgements}
We thank Hy Trac for helpful discussions. 
NC acknowledges support from the McWilliams Graduate Fellowship.
NC and DM acknowledge support from the McWilliams Center-Pittsburgh Supercomputing Center seed grant.
DM also acknowledges support from NASA grant 80NSSC22K0722.
TDM and RACC acknowledge funding from 
the NSF AI Institute: Physics of the Future, NSF PHY-2020295, 
NASA ATP NNX17AK56G, and NASA ATP 80NSSC18K101. 
TDM acknowledges additional support from  NSF ACI-1614853, NSF AST-1616168, NASA ATP 19-ATP19-0084, NASA ATP 80NSSC20K0519, and RACC from NSF AST-1909193.
YN acknowledges support from the ITC Postdoctoral Fellowship.
SB acknowledges funding from NASA ATP 80NSSC22K1897.

%%%%%%%%%%%%%%%%%%%%%%%%%%%%%%%%%%%%%%%%%%%%%%%%%%
\section*{Data Availability}

The code to reproduce the simulation is available at \url{https://github.com/MP-Gadget/MP-Gadget}, and continues to be developed. Text file forms of the data presented here and scripts to generate the figures are available. 
The resimulation initial conditions and snapshots are available upon reasonable request to the authors.

\bibliographystyle{mnras}
\bibliography{main.bib}

\appendix
\input{AppA_convergence}

\end{document}

%% file: Sec1_Intro.tex
\section{Introduction}
Observations of local galaxies suggest that a supermassive black hole (SMBH) is harbored in almost all galactic centers \citep[e.g.][]{Tremaine2002, Kormendy2013}.
These SMBHs has already grown to $\gtrsim 10^7\,M_\odot$ and some even to $\gtrsim 10^9\,M_\odot$ at high redshift ($z\sim 6$) through observations of high-redshift quasars \citep[e.g.][]{Fan2001, Banados2018, Wu2015, Wang2021}.
They are thought to have formed in the high-redshift Universe ($z\sim 20$), but the exact seeding mechanism remains largely unconstrained \citep[e.g.][]{Woods2019} due to their low masses and faint electromagnetic emissions \citep[e.g.][]{Reines2016}.

Recently, an MBH was found at $z\sim 11$ \citep{Maiolino2023}, and more high-redshift MBHs have been revealed by JWST \citep[e.g.][]{Ubler2023,Kocevski2023, Harikane2023, Matthee2023}.
These MBHs are found to be over-massive compared to their host galaxies compared with the low-redshift relation \citep[e.g.][]{Pacucci2023}.
Such over-massive MBHs pose new challenges to the growth of MBHs in early galaxies, especially for MBH seeding by Pop-III stars, or runaway stellar growth in dense star clusters, since they usually need to grow at super-Eddington rates to reach the mass of the observed high-redshift MBHs.

Gravitational waves (GWs) from MBH mergers offer a promising way to observe the first MBH seeds \citep[e.g.][]{Sesana2005, Barausse2012, Klein2016, Ricarte2018}, especially when combined with observations of the electromagnetic (EM) counterparts \citep[][]{Natarajan2017,DeGraf2020}.
The gravitational waves of MBH mergers with masses in the range $10^4-10^7\, M_\odot$ have a frequency around mHz, and they are primary targets for the Laser Interferometer Space Antenna (LISA), which can detect MBHs with such masses out to $z>20$ \citep[][]{LISA2017arXiv170200786A}.
Compared to electromagnetic observations, GWs not only allow us to probe MBH seeds at higher redshift, but also provide MBH mass estimations independent of their instantaneous accretion state. 
However, modeling of MBH mergers depends heavily on the dynamics during the formation and shrinking of black hole (BH) binaries which are poorly constrained from kpc \citep[e.g.][]{Tremmel2017, Pfister2019}, to pc \citep[e.g.][]{Colpi2014} scales.
This can lead to a large spread in detection rates for LISA depending on the assumptions made \citep[e.g.][]{Sesana2010, Klein2016}.
Therefore, an accurate understanding of the dynamical journey of seed MBHs in the early galaxy assembly is key to robust constraints on the seed MBH population with GW detections.

The dynamics of MBH pairs towards coalescence are first summarized in \cite{Begelman1980}.
During galaxy mergers, the central MBHs start at a large separation in the remnant galaxy (a few tens of kpc). 
The MBHs then gradually lose their orbital energy and sink to the center of the remnant galaxy due to the dynamical friction exerted by the gas, stars, and dark matter around them \citep[e.g.][]{Chandrasekhar1943,Ostriker1999}. 
When their separation is $\lesssim 1$ parsec, an MBH binary forms, and other energy-loss channels begin to dominate, such as scattering with stars \citep[e.g.][]{Quinlan1996,Berczik2006,Sesana2007b,Berentzen2009,Khan2011,Khan2013,Vasiliev2015}, gas drag from the circumbinary disk \citep[e.g.][]{Haiman2009, Lai2023},  and three-body scattering with a third black hole \citep[e.g.][]{Bonetti2018, Mannerkoski2021}.

Currently, our most accurate understanding of the MBH binary population and its relation with galaxy evolution comes from cosmological simulations \citep[see e.g.][ for an overview]{lisa_astro2023}.
Cosmological simulations self-consistently follow the coevolution of MBHs with host galaxies, and contain rich information about the environments where MBH interactions and mergers take place  \citep[e.g.][]{Salcido2016, Kelley2017b, Tremmel2017, Volonteri2020, Katz2020, Chen2022}.
These environments include a wide range of scenarios from isolated dwarf galaxies to infalling satellites of a massive central galaxy in a cluster.
However, realistic and large-volume modeling comes at the cost of limited resolutions, and at best they can follow MBH dynamics to $\sim {\rm kpc}$ scales.
These simulations are also subject to the simplistic subgrid-seeding mechanism that only considers the higher end of the MBH seed mass \citep[out of these, ][probes relatively low MBH seeds at high-redshifts]{Dubois2014, Tremmel2015}.
To compensate for the resolution limit in large-volume simulations, cosmological ``zoom-in" simulations are also applied to study the sinking behavior of specific merging systems \citep[e.g.][]{Pfister2019, Bortolas2020}, but they are also subjected to high-computational cost and low flexibilities in configuration and subgrid models.
Only a few galaxies and the MBHs can be studied at a time.

To accurately model the MBH orbital evolution on sub-kpc scales,  high-resolution, idealized galaxy merger simulations, and direct N-body simulations are often used to investigate the detailed dynamical processes of galaxy/MBH mergers \citep[e.g.][]{Khan2016a, Gualandris2022,  Liao2023a}.
These methods have great flexibility in varying the galaxy properties, orbital configurations, and subgrid models, and allow us to gain a detailed understanding of how different physics mechanisms impact the orbital decay and hardening of the MBH pairs and binaries.
Most of the idealized merger simulations, however, do not account for the fully realistic scenario of the orbital properties of the MBH pairing and consecutive galaxy mergers frequent at high redshifts.

Very recently, many emerging works have started to consider the more realistic scenarios of MBH dynamics.
This realism is approached in various ways.
For example, \cite{Mannerkoski2021} and \cite{Koehn2023} directly recover the initial condition of cosmological mergers with accurate, high-resolution N-body methods and study the dynamics of binary and triple SMBHs across a wide dynamical range.
\cite{Partmann2023} considers the scenario of multiple infalling satellites with seed MBH, which is typical for a high-redshift massive galaxy, and studies the many-body interactions with a treatment of gravitational recoils.
\cite{Liao2023a, Liao2023b} used realistic subgrid models of hydrodynamics simulations in combination with small-scale treatment of binary hardening to study the impact of galaxy types and physics modeling on the merging timescale of SMBH binaries.

To truly bridge the gap between cosmological simulations and small-scale MBH dynamics, we introduce the ``Massive Black Hole Assembly in Galaxies Informed by Cosmological Simulations" (MAGICS) suite, which combines the realistic MBH and galaxy population from the state-of-the-art cosmological hydrodynamics simulation \texttt{ASTRID} with the idealized galaxy merger simulations.
This suite (also referred to as ``resimulations" hereafter) directly recovers the high-redshift galaxy merger properties to include both isolated and multiple galaxy mergers, and uses the ``full-physics" hydrodynamical subgrid modeling with star formation and various feedback channels.
This is the first work to recover not only the collisionless component (e.g. dark matter and stars) in cosmological MBH merger events but also the full hydrodynamics evolution of the merging system.
The importance of the latter in gas-rich environments (typical of high-redshift galaxies) has already been pointed out in e.g. \cite{Fiacconi2013}, \cite{Tamburello2017}, \cite{Liao2023b}.

This paper is organized as follows.
Section \ref{sec:sim} introduces the simulation code and the subgrid models used in the simulations.
Section \ref{sec:astrid_pop} gives an overview of the high-redshift MBH merger population in \texttt{ASTRID}, which is the base population for the high-resolution resimulation suite.
We also show the detailed properties of the systems selected for the first suite of high-resolution resimulations in this work.
In Section \ref{sec:multigal}, we describe how we set up galaxy merger initial conditions to directly mimic the original cosmological system.
Finally, in Section \ref{sec:result} we present the results for the dynamical friction time scales of seed MBH mergers in various environments, and the correlation with large-scale galaxy and orbital properties.
We also investigate the effect of multiple galaxy mergers on the MBH sinking and the inclusion of nuclear star clusters.

% papers to cite:
% resimulation: \citep[][]{Khan2016a, Koehn2023}
% high-z mergers: \citep[][]{Pfister2019, Tamburello2017, }

%% file: Sec2_Method.tex
\section{Idealized Galaxy Simulation with MP-Gadget}
\label{sec:sim}

  \begin{figure*}
    \centering
    \includegraphics[width=0.89\textwidth]{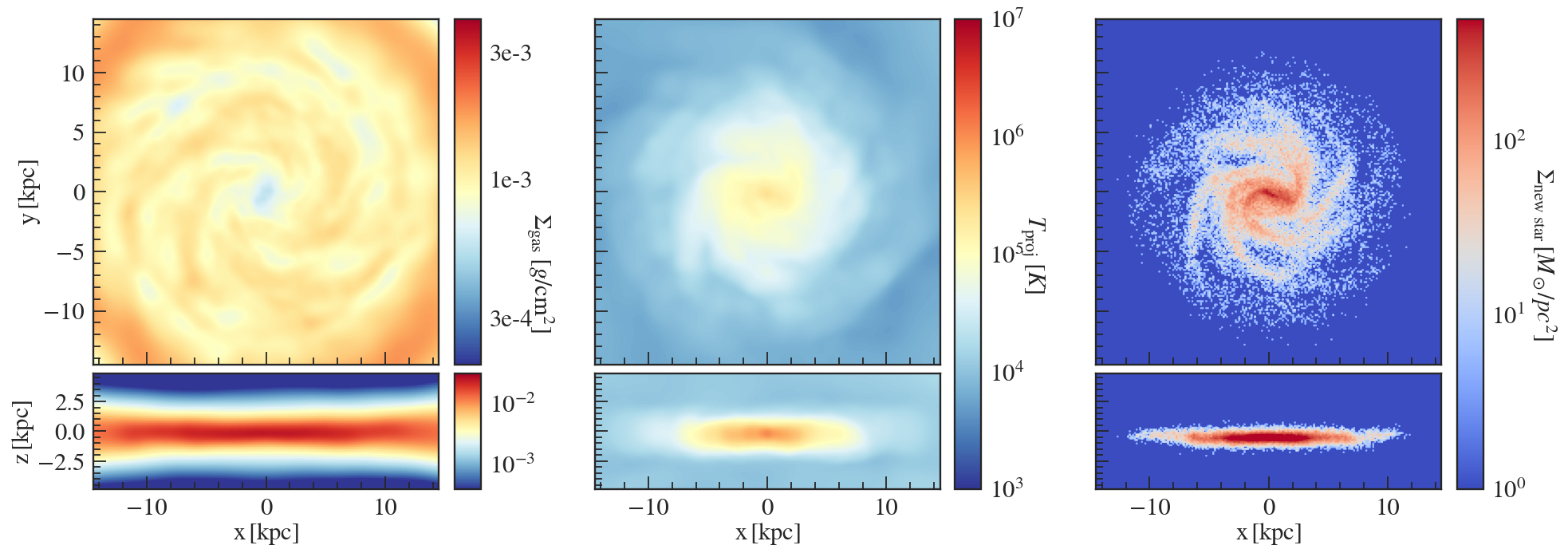}
    \includegraphics[width=0.89\textwidth]{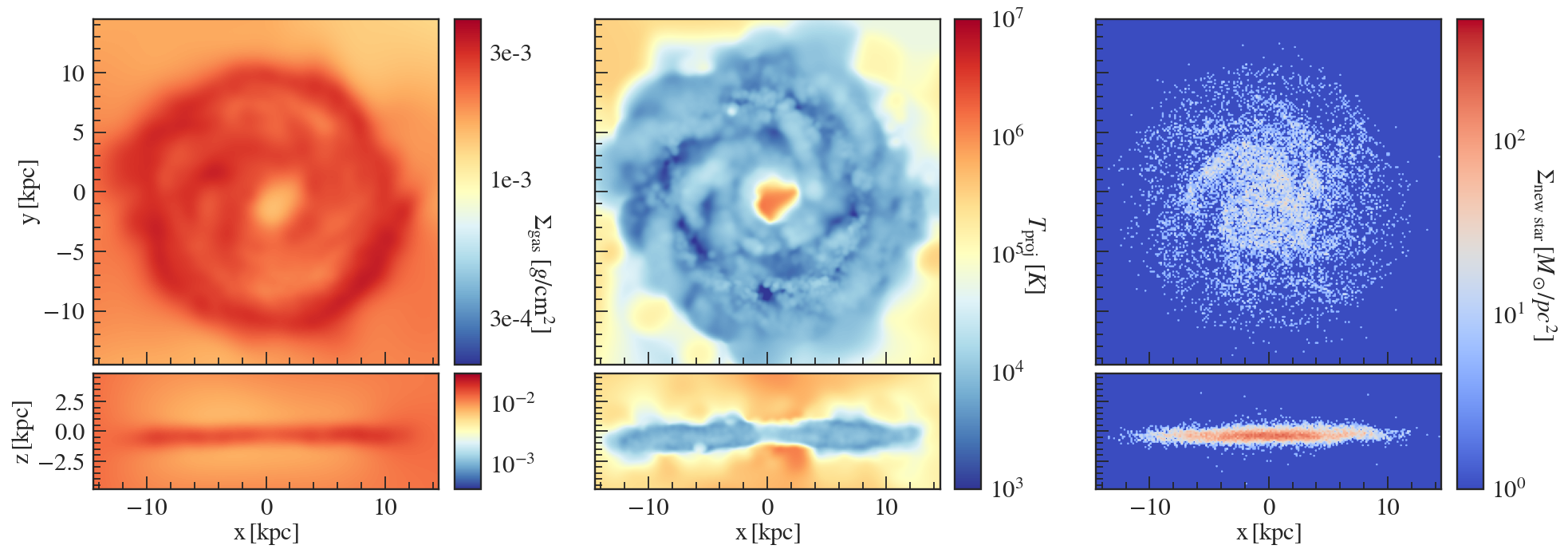}
    \includegraphics[width=0.89\textwidth]{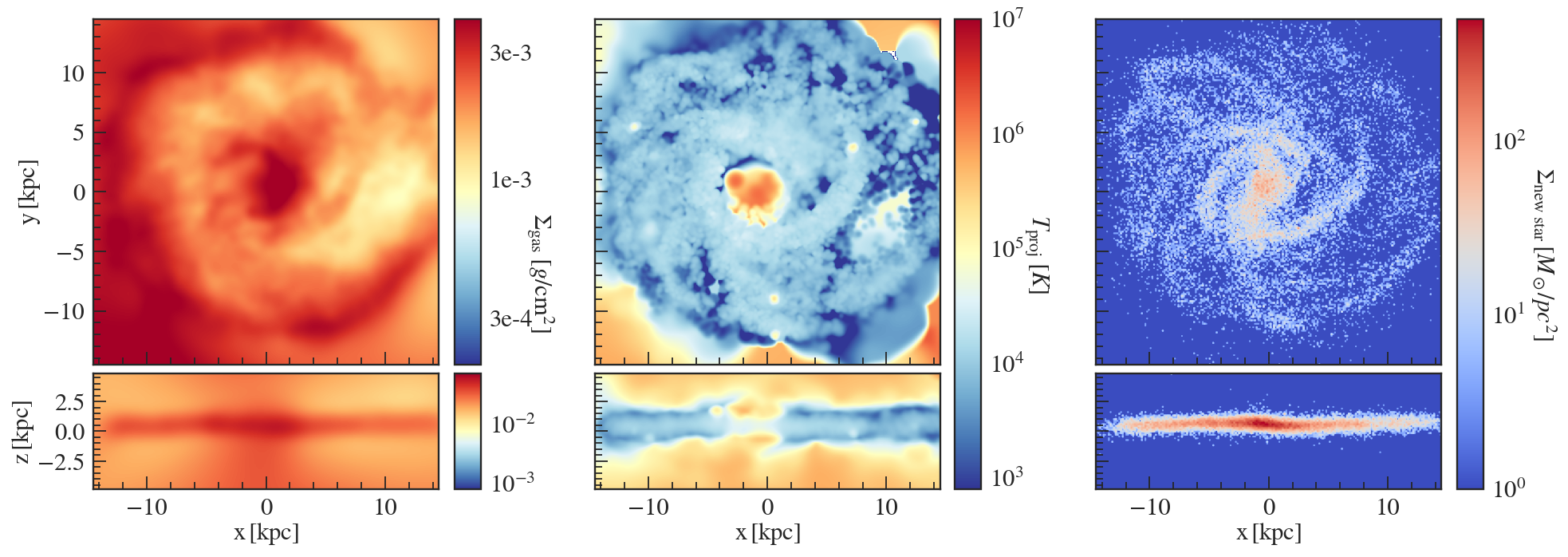}
    \caption{\textit{Top Row:} Evolution of a disk galaxy with the ``SH03" model in \texttt{MP-Gadget} after 500 Myrs. 2D projected gas density (\textit{left}), density-weighted temperature (\textit{middle}) and 2D projected density of newly-formed stars (\textit{right}). \textit{Middle Row:} Evolution of the same disk galaxy run with the \texttt{ASTRID} subgrid model in \texttt{MP-Gadget} after 500 Myrs. \textit{Bottom Row:} evolution of the same disk galaxy as the middle row  run with the \texttt{ASTRID} subgrid model, but with $50\%$ of the gas put into the gas halo component. }
    \label{fig:val_prof2d_sfcool}
\end{figure*}

\subsection{The subgrid physics model for galaxy formation}
The subgrid gas, black hole, and galaxy-formation physics in the resimulations largely follow from the model in the \astrid cosmological simulation \citep{Bird2022, Ni2022}.
We summarize the key components here.
In our simulations, gas cools via primordial radiative cooling \citep{Katz1996} and via metal line cooling, with the gas and stellar metallicities traced following \cite{Vogelsberger2014}.
In the context of isolated-galaxy simulation, we do not include the patchy reionization model.
The ionizing ultra-violet background from \cite{FG2020} is employed with gas self-shielding being factored in as outlined in \cite{Rahmati2013}.
Star formation is based on a multi-phase model for stellar formation as described in \cite{SH03}, accounting for the influence of molecular hydrogen~\citep{Krumholz2011}.
Type II supernova wind feedback is incorporated into the simulation in accordance with \cite{Okamoto2010}, with wind speeds proportional to the local one-dimensional dark matter velocity dispersion.

% Subgrid models for MBHs are summarized as follows.
MBHs are represented by particles that can accrete gas, merge, and apply feedback to the surrounding gas medium.
For this work, we do not seed extra BHs during the resimulation, but include them in the initial conditions.
Gas accretion onto BHs is modeled with a Bondi-Hoyle-Lyttleton-like prescription \citep{DSH2005}:
\begin{equation}
\centering
\label{equation:Bondi}
    \dot{M}_{\rm B} = \frac{4 \pi \alpha G^2 M_{\rm BH}^2 \rho}{(c^2_s+v_{\rm rel}^2)^{3/2}}
\end{equation}
where $c_s$ and $\rho$ are the local sound speed and density of the gas, $v_{\rm rel}$ is the relative velocity of the BH with respect to the nearby gas, and $\alpha = 100$ is a dimensionless fudge parameter to account for the underestimation of the accretion rate due to the unresolved cold and hot phase of the subgrid interstellar medium in the surrounding.
We allow for short periods of super-Eddington accretion in the simulation but limit the accretion rate to two times the Eddington accretion rate.
The BH radiates with a bolometric luminosity $L_{\rm bol}$ proportional to the accretion rate $\dot{M}_\bullet$, with a mass-to-energy conversion efficiency $\eta=0.1$ in an accretion disk according to \cite{Shakura1973}.
\begin{equation}
\centering
\label{equation:Lbol}
    L_{\rm Bol} = \eta \dot{M}_{\rm BH} c^2
\end{equation}
$5\%$ of the radiated energy is coupled to the surrounding gas as the AGN feedback.

The dynamics of the BHs are modeled with a sub-grid dynamical friction model \citep{Tremmel2015,Chen2021} in both \texttt{ASTRID} and the resimulations.
This model provides an improved treatment for calculating BH trajectories and velocities.
Two BHs merge if their separation is within two times the spatial resolution $2\epsilon_{\rm g, BH}$ (this is $\sim 500\,{\rm pc}$ in \texttt{ASTRID} at $z=6$, and $20\,{\rm pc}$ in the resimulations), once their kinetic energy is dissipated by dynamical friction, and they are gravitationally bound to the local potential.
We note that since we numerically merge the MBHs at $20\,{\rm pc}$ in the resimulations, our modeling ends before the MBH pairs become a bound binary. Therefore we do not attempt to model binary formation and binary hardening process (except in Section \ref{sec:taichi}).
In the \texttt{ASTRID} simulation, we use a separate mass tracer $M_{\rm dyn}$
to reduce the noisy gravitational forces (dynamical heating) acting on the small seed mass black holes during the force calculations of BHs (gravity and dynamical friction). 
When a new BH is seeded, $M_{\rm dyn}$ is set to $M_{\rm dyn,seed} = 10^7 \hmsun$, which is about $1.5 M_{\rm DM}$.
$M_{\rm dyn}$ is kept at its seeding value until $M_{\rm BH}>M_{\rm dyn,seed}$. After that, $M_{\rm dyn}$ grows following the BH mass accretion.

Although this approach is a necessary step to alleviate dynamic heating and stabilize the BH motion in the early growth phase, it can also lead to underestimation of the DF timescale and over-predict high-redshift seed MBH mergers.
In the high-resolution resimulations, with a stellar particle mass of $2000\msun$, we alleviate the boost of the dynamical mass of BH particles in the original simulation and use $M_{\rm dyn}=2\times 10^5\,M_\odot$.
This is $\sim 100$ times smaller than the values used in \texttt{ASTRID} and gives a more faithful estimation of the true merging timescale.
In future works, we will further push the resolution limit to directly model the true mass of seed MBHs.
This typically requires a mass ratio of $> 10$ between dark matter and MBH particles with moderate softening and the inclusion of DF subgrid modeling, and $\sim 1000$ without  \citep{Pfister2019, Ma2021}.

\subsection{Code Validation}

\begin{figure}
    \centering
    \includegraphics[width=0.48\textwidth]{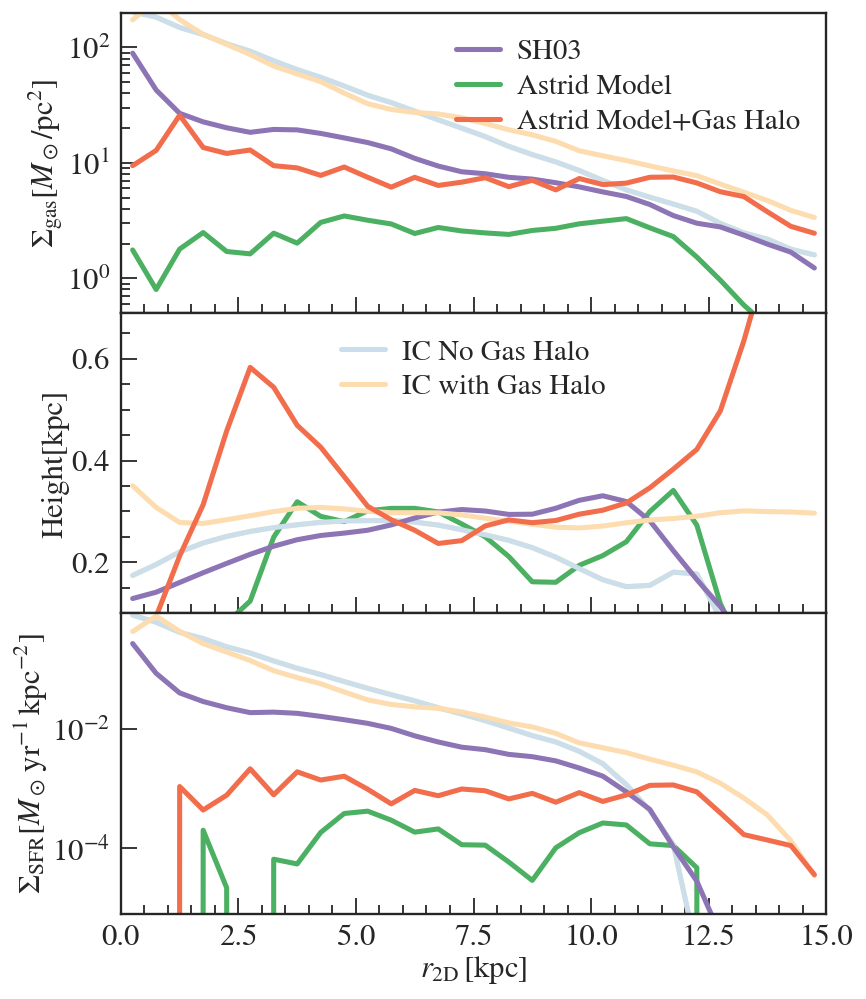}
    \caption{Comparison between the gas surface density profile (\textit{top panel}), gas disk height (\textit{middle panel}), and star-formation rate (SFR) surface density between the three runs. The lines with light colors show the gas properties in the initial conditions, and the lines with dark colors show the properties after 1 Gyr of evolution. The ``SH03" run (purple) and the ``Astrid Model" run (green) share the same IC (light blue). The ``Astrid Model+Gas Halo" run (orange) has $50\%$ gas in the disk and $50\%$ gas in the halo for the initial condition (light orange).}
    \label{fig:val_gas_prof}
\end{figure}

\begin{figure}
    \centering
    \includegraphics[width=0.46\textwidth]{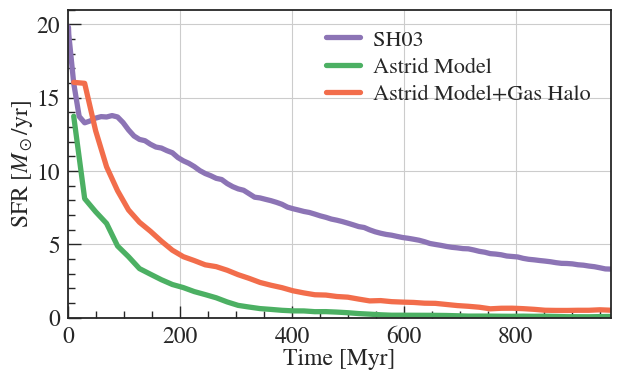}
    \caption{Evolution of the total star formation rate in the three validation runs. Without wind and AGN feedback, the ``SH03" run has the most (up to $\sim 4$ times higher) star formation throughout the simulation.}
    \label{fig:val_sfrh}
\end{figure}

 \subsubsection{Disk Galaxy with Star Formation}
 % \nianyi{ add future improvements, specifically ISM modeling and possible effects on BH sinking etc}
 
We first validate the physical models and implementations in \texttt{MP-Gadget} in the context of an isolated disk galaxy.
We set up the initial condition (IC) following the Agora code comparison suite \citep{agora1, agora2}, using the \texttt{MakeDisk} IC generator \cite{Springel2005b}, with parameters matching with Table 3 of \cite{agora2}, except that we use a disk gas fraction of 0.4 to match our high-redshift application.
We model the multiphase interstellar medium following the prescription of \cite{SH03} that incorporates gas cooling, star formation, and SN thermal feedback. (this run is denoted as ``SH03"). 
%We also enable metal line cooling, with an initial constant gas metallicity of $0.01$. 
In this first validation step, we did not match the full \texttt{ASTRID} physical models: we did not include the wind feedback from TypeII supernova and the influence of molecular hydrogen on star-formation; we also did not include BHs in this run, so there is no BH accretion and feedback.
We run the simulation for $1\,{\rm Gyr}$ and observe a stable disk throughout the simulation.

In the top row of Figure \ref{fig:val_prof2d_sfcool}, we show the 2D projected density and density-weighted temperature of the gas, along with the projected density of newly formed stars after 500 Myrs of evolution.
In this model, the central star formation is high because we do not include feedback mechanisms to mitigate the gas condensation.
In Figure \ref{fig:val_gas_prof}, we show the 2D profiles of gas properties in the initial condition and after 1 Gyr of evolution for this run (light blue and purple lines).
The gas surface density is computed as the total gas mass in each radial bin divided by the area of the bin.
The disk height is the mass-weighted distance to the x-y plane for particles in each radial bin.
We note that to compare with runs with feedback and a gas halo, where the disk component is only a fraction of the total particle, we only take the star-forming gas with $|z| < 2\,{\rm kpc}$ in the disk height computation.
Finally, the star-formation rate (SFR) surface density is the total SFR in each radial bin divided by the area of the bin.
The gas density and SFR decrease throughout 1 Gyr due to the gradual depletion of gas, while we can maintain a thin disk throughout the evolution.

Figure \ref{fig:val_sfrh} shows the total SFR in the isolated galaxy over 1 Gyr.
Since our initial gas fraction in the disk is two times larger than the Agora suite, the SFR is also higher.
After 500 Myrs of evolution, $18\%$ of the stars are newly formed out of $26\%$ of disk gas, mostly residing at the center of the galaxy, as can be seen from the top-right panel of Figure \ref{fig:val_prof2d_sfcool}.

 \subsubsection{Disk Galaxy with ASTRID Models}
To match the physical models used in the \texttt{ASTRID} simulation, as a further validation of the resimulation subgrid modeling we include a BH with an initial mass of $4\times 10^6\, M_\odot$ and turn on all the \texttt{ASTRID} subgrid physics models described in Section \ref{sec:sim}.
Compared with the vanilla SH03 modeling shown in the previous section, adding wind and AGN feedback is expected to remove the dense gas at the galaxy center and regulate the star formation in the disk \citep[e.g.][]{DiMatteo2005, Okamoto2010, Weinberger2017}.
We use two different initial conditions for this ``Astrid-Model" run: we first keep the same IC as the ``SH03" run in the previous section, with only the addition of a central BH.
Then, to match more closely with the high-redshift gas environments where only a small fraction of the total gas in the halo is star-forming and in disk structure ($\sim 10-40\%$, see e.g. Table \ref{tab:ic} in Appendix \ref{app:ic}), we use a modified version of \texttt{MakeDisk} to put $50\%$ of the total gas into the gaseous halo component, following the method laid out in \cite{Su2019}.
We set up the gas halo in thermal equilibrium following a $\beta$ profile with $\beta=0.66$ and $R_c/R_s = 0.5$.
We adiabatically relax the IC for $250\,{\rm Myrs}$ until it becomes stable, before turning on other subgrid models.

In Figure \ref{fig:val_prof2d_sfcool} we show comparisons of projected gas properties and newly formed stars in the SH03 run, the ``Astrid Model" run with all gas in the disk, and the ``Astrid Model" run with a gas halo (``Astrid Model+Gas Halo").
With feedback models turned on, we see a significant drop in central star formation (the right column), and the disk gas can cool to much lower temperatures (middle column).
We find that the supernova wind feedback is more efficient at removing gas from the galaxy center and thus lowering the central SFR, compared with AGN feedback.
Figure \ref{fig:val_gas_prof} shows the comparisons between the gas profiles in the three runs, along with their corresponding ICs (note that the ``SH03" run and the ``Astrid Model" run share the same IC so it is only shown once by the light blue curve).
We see that the disk height and a high star-formation rate are maintained throughout the ``SH03" run.
The ``Astrid Model+Gas Halo" run has a flattened star-formation surface density and gas surface density after a Gyr of evolution.

Finally, Figure \ref{fig:val_sfrh} shows the total SFR in the isolated galaxies in the three runs during 1 Gyr.
The ``SH03" run has the most (up to $\sim 4$ times higher) star formation throughout the simulation. 
The model with a gaseous halo displays more efficient star formation compared to the no-halo run because more central gas is pushed to further distances by wind feedback in the disk-only model.
In the ``Astrid Model+Gas Halo" run we observe a continuous inflow of cold gas clumps, which sustains the star formation in the disk for a longer time.
These clumps can also impact the dynamics of seed MBHs.
We will use the ``Astrid Model+Gas Halo" modeling in all of the resimulations in this work.

%% file: Sec3_Population.tex
\section{High-Redshift MBH Mergers}
\label{sec:astrid_pop}

\begin{figure*}
    \centering
    \includegraphics[width=0.92\textwidth]{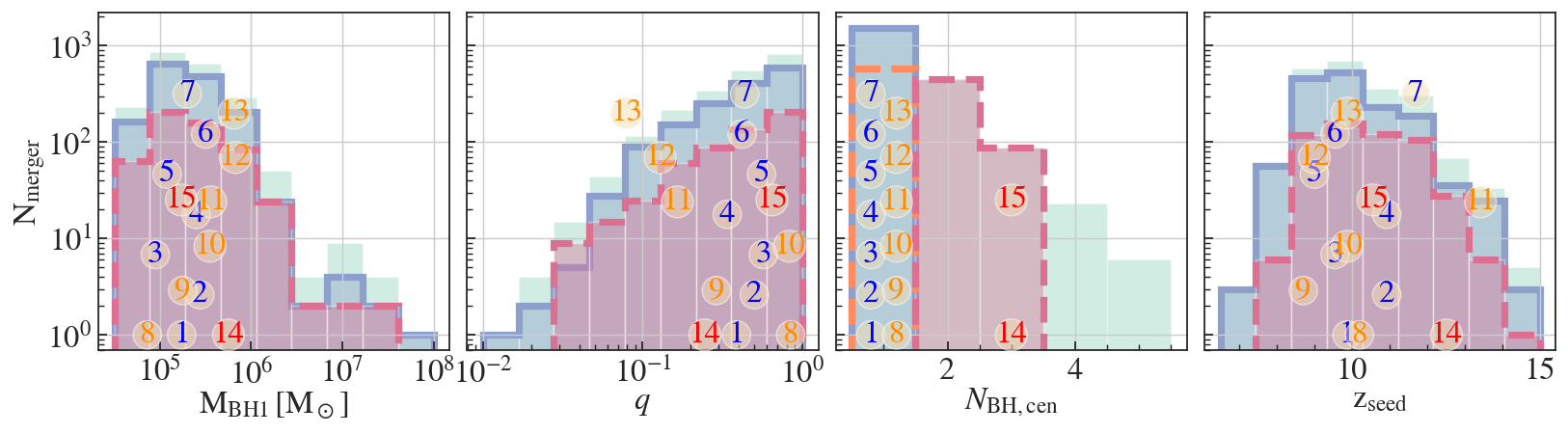}
    \caption{The $z=6$ MBH merger population in \texttt{Astrid}. From left to right: the mass distribution of the more massive MBH among the pair (\textbf{first panel}); the distribution of the mass ratio between the two merging MBHs (\textbf{second panel});  the number of MBHs in the central region ($<3{\rm kpc}$ from the galaxy center) of the remnant galaxy (\textbf{third panel}); the seeding redshift of the more massive BH in the pair (\textbf{fourth panel}). The background \textbf{green} histogram shows the entire merger population, and we also show two sub-populations: the \textbf{blue} histogram is the "isolated" mergers with no third MBH coming into the central region of the host galaxy before $z=6$, and the \textbf{red} histogram shows the "complex" mergers with multiple MBH in the host galaxy center already at $z=6$. The numbers overlaid on each plot label where the resimulated systems lie within each distribution (only the x-values are meaningful, and the y-values are randomly taken).}
    \label{fig:z6_pop_bh}
\end{figure*}

\begin{figure*}
    \centering
        \includegraphics[width=0.99\textwidth]{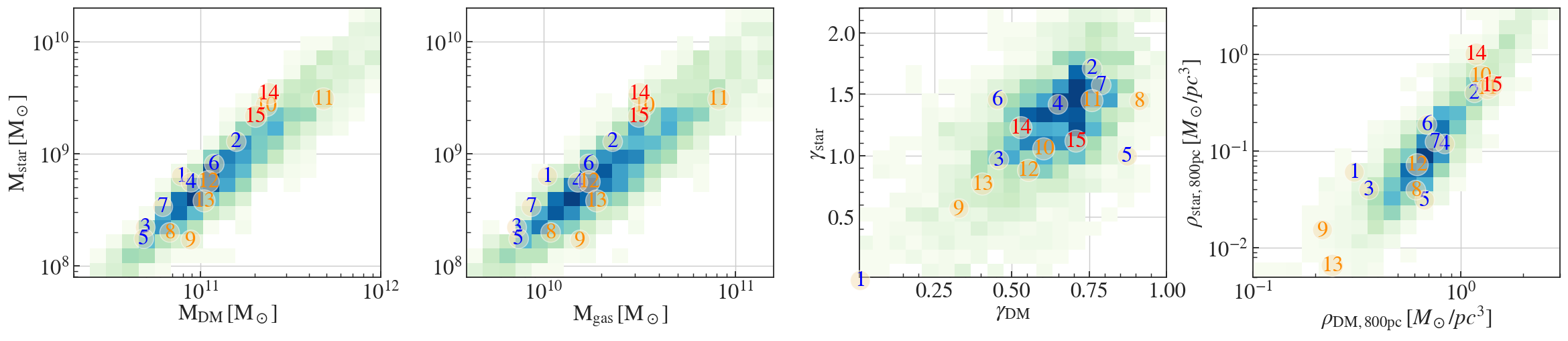}
    \caption{Host galaxy properties of the $z=6$ MBH merger remnant in \texttt{Astrid}. \textit{First and second panels:} the 2D distribution of the galaxy mass with the dark matter halo and total gas mass in the halo. \textit{Third panel:} power-law index of the dark matter and stellar density profiles measured at the \texttt{ASTRID} resolution. \textit{Fourth panel:} dark matter and stellar densities measured at at the \texttt{ASTRID} resolution ($0.8\,{\rm kpc}$ from the galaxy center).}
    \label{fig:z6_pop_gal}
\end{figure*}

\begin{figure}
    \centering
    \includegraphics[width=0.42\textwidth]{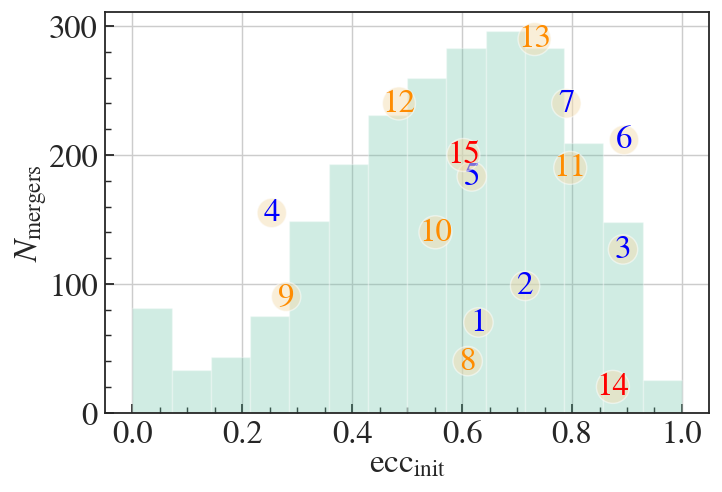}
    \caption{Initial orbital eccentricities of the merging MBH pair in \texttt{ASTRID}, calculated from the first periapsis and the first apoapsis. The overall distribution peaks at $\sim 0.7$, and the selected resimulation systems (scattered numbers) cover a range of eccentricities from $0.2$ to $0.9$ (similar to Figure \ref{fig:z6_pop_bh}, the y-values are randomly chosen for readability).}
    \label{fig:z6_pop_orb}
\end{figure}

% One key science goal of LISA is to address the origin and growth of MBHs and answer questions on how MBHs form and evolve in the early Universe, and how they assemble with time and become present in almost all the galaxies in the local Universe.
Previous works have shown that the dynamical friction timescales are long for MBH seeds, but with MBH seeds modeled down to $5\times 10^4\,{M_\odot}$, we should still expect order unity mergers per year, dominated by seed-seed mergers \citep[e.g.][]{Tremmel2017, Volonteri2020, Chen2022}.
However, the seed-seed merger is also the regime where the current simulation lacks prediction confidence due to the limited particle resolution.
In this section, we examine the high-z merger population in the \texttt{ASTRID} simulation and select typical cases from the population for comparison with high-resolution resimulation.

\subsection{$z\sim 6$ merger population in ASTRID}

We focus on resimulating the population of $z\sim 6$ merger events in the \texttt{ASTRID} simulation, to study the sub-kpc scale sinking and merger timescales of seed MBHs.
Here we give a brief overview of this population and their environments.
There are 2107 MBH mergers between $z=6-6.2$ in \texttt{ASTRID}.
We collect their host galaxy properties from the $z=6$ snapshot (i.e. shortly after the merger).

In Figure \ref{fig:z6_pop_bh}, we show the MBH properties of all mergers in this redshift bin, including the primary mass, mass ratio, the total number of MBHs in the remnant galaxy, and the seeding redshift of the primary MBH.
We subdivide the merger population into three groups, according to the total number of MBHs in the remnant galaxy at $z=6$ (we take this as a proxy for the complexity of the host environment).
The first group is the ``isolated mergers" which have no new infalling structures interacting with the merging MBHs at least until $z=6$ (blue population in Figure \ref{fig:z6_pop_bh}).
This group consisted of the majority of high-z mergers (1531/2107), with the merger remnant being the only MBH in its host galaxy.
Within these ``isolated mergers", however, a significant fraction will have new infalling MBHs/galaxies soon after the \texttt{ASTRID} merger at $z=6$ (i.e. we found new structures on the outskirts of the remnant halo at $z=6$).
Although these new infalls will not impact MBHs that already merged in \texttt{ASTRID} by $z=6$, they may interfere with the binary formation in the high-resolution resimulations. 
Hence, we treat these systems separately (shown as the orange bar in the third panel of Figure \ref{fig:z6_pop_bh}) when setting up resimulations.

%%%%%%%%%%%%%%%%%%%%%%%%%%%%%%%%%
\begin{table*}
\centering
\caption{Properties of the \texttt{ASTRID} galaxy/MBH merger systems selected for high-resolution resimulations (Also see Figures \ref{fig:z6_pop_bh}, \ref{fig:z6_pop_gal}, \ref{fig:z6_pop_orb}). Column 1 is the label of each system used throughout the paper. Columns 2-4 are the two MBH and host properties measured before the galaxy merger (at $z_{\rm init}$). Column 5 is the eccentricity of the first orbit in \texttt{ASTRID}. Column 6 is the redshift when we initialized the resimulations. Columns 7-10 are the host properties of the merger remnant at $z=6$. The last column is the number of MBHs in the remnant galaxy. The horizontal lines divide mergers in isolation until at least $z=4$ (systems 1-7), mergers in isolation from $z=9-6$ but with new infalling galaxies soon after $z=6$ (systems 8-13), and mergers between multiple galaxies and MBHs (systems 14-15).}
\label{tab:astrid}
\begin{tabular}{lcccccccccc}

\hline
Name & $M_{\rm BH\,1,2}$  & $M_{\rm halo\,1, 2}$ & $M_{\rm gal\,1, 2}$ & $e_{\rm init}$ &  $z_{\rm init}$ & ${\rm M}_{\rm halo, rem}$ & $M_{\rm gas, rem}$ & $M_{\rm gal, rem}$ & ${\rm SFR}_{\rm rem}$ & $N_{\rm BHs, rem}$ \\
& [$10^5\,h^{-1} {\rm M}_\odot$]  & [$h^{-1} {\rm M}_\odot$]  & [$h^{-1} {\rm M}_\odot$]  &   &  & [$h^{-1} {\rm M}_\odot$] &  [$h^{-1} {\rm M}_\odot$] & [$h^{-1} {\rm M}_\odot$] & [$M_\odot/{\rm yr}$] & \\
\hline
\texttt{system1} & 1.7, 0.7 & 1e10, 2e10 & 8e6, 4e7 & 0.63 & 9 & 7e10 & 1e10 & 6e8 & 3.0 & 1\\
\texttt{system2} & 2.7, 1.4 & 4e10, 4e10 & 2e8, 6e7 & 0.71 & 7.6 & 1e11 & 2e10 & 1e9 & 9.1 & 1\\
\texttt{system3} & 0.9, 0.5 & 1e10, 9e9 & 9e6, 1e7 & 0.90 & 9 & 5e10 & 7e9 & 2e8 & 0.86 & 1\\
\texttt{system4} & 2.5, 0.8 & 3e10, 7e9 & 4e7, 2e7 & 0.26 & 9 & 9e10 & 2e10 & 5e8 & 7.8 & 1\\
\texttt{system5} & 1.2, 0.7 & 2e10, 5e9 & 1e7, 1e7 & 0.60 & 7.6 & 5e10 & 7e9 & 2e8 & 1.1 & 1\\
\texttt{system6} & 3.2, 1.3 & 2e10, 7e9 & 1e7, 8e6 & 0.89 & 9 & 1e11 & 2e10 & 8e8 & 6.2 & 1\\
\texttt{system7} & 2.0, 0.9 & 2e10, 7e9 & 2e7, 7e6 & 0.79 & 9 & 6e10 & 9e9 & 3e8 & 1.1 & 1\\
\hline
\texttt{system8} & 0.7, 0.6 & 2e10, 2e9 & 2e7, 1e7 & 0.61 & 7.6 & 7e10 & 1e10 & 2e8 & 3.3 & 1\\
\texttt{system9} & 1.8, 0.5 & 1e10, 1e10 & 3e6, 7e6 & 0.29 & 7.6 & 9e10 & 2e10 & 2e8 & 1.6 & 1\\
\texttt{system10} & 3.5, 3.0 & 3e10, 3e10 & 3e7, 3e7 & 0.55 & 9 & 2e11 & 3e10 & 2e9 & 15 & 1\\
\texttt{system11} & 3.6, 0.6 & 1e11, 1e10 & 2e8, 3e7 & 0.80 & 7.6 & 5e11 & 8e10 & 3e9 & 53 & 1\\
\texttt{system12} & 6.7, 0.9 & 2e10, 1e10 & 2e7, 2e7 & 0.46 & 7.6 & 1e11 & 2e10 & 5e8 & 3.2 & 1\\
\texttt{system13} & 6.5, 0.5 & 1e10, 3e9 & 1e7, 1e7 & 0.74 & 9 & 1e11 & 2e10 & 4e8 & 3.5 & 1\\
\hline
\texttt{system14} & 5.6, 1.4 & 3e10, 6e10 & 8e7, 2e8 & 0.88 & 9 & 2e11 & 3e10 & 4e9 & 13 & 3\\
\texttt{system15} & 1.7, 1.1 & 8e10, 4e9 & 2e8, 2e7 & 0.59 & 9 & 2e11 & 3e10 & 1e9 & 10 & 3\\

\hline
\end{tabular}
\end{table*}
%%%%%%%%%%%%%%%%%%%%%%%%%%%%%%%%%
\begin{table}
\caption{Mass and spatial resolutions of the resimulation suite. The maximum separation for two MBHs to merge in the simulation is $2\times \epsilon_{\rm BH}$.}
\setlength\tabcolsep{1pt}
\begin{tabular*}{\columnwidth}{@{\extracolsep{\fill}}rccccccr}
\toprule
$M_{\rm DM}$ & $M_{\rm gas}$  & $M_{\rm star}$ & $M_{\rm BH, dyn}$ & $\epsilon_{\rm DM}$ &  $\epsilon_{\rm gas}$ & $\epsilon_{\rm star}$ & $\epsilon_{\rm BH}$\\
\hline
8000 $M_\odot$ & 8000 $M_\odot$ & 2000 $M_\odot$ & $2\times 10^5\,M_\odot$ & $80\,{\rm pc}$ & $80\,{\rm pc}$ &$20\,{\rm pc}$ & $10\,{\rm pc}$ \\
\hline
\end{tabular*}
\label{t:stats}
\end{table}

%%%%%%%%%%%%%%%%%%%%%%%%%%%%%%%%%%

The second group consists of ``multiple MBH" systems (the red population), where at $z=6$, the galaxy remnant has $1-2$ other MBHs besides the merging pair.
This means that the orbits of the pair go through more complex interactions already before $z=6$ with other galaxies/MBHs. 
From Figure \ref{fig:z6_pop_bh}, we see that about $25\%$ (542 out of 2107) of $z=6$ MBH pairs reside in these multiple galaxy interaction environments. 
This highlights the importance of considering multiple MBH interactions when modeling the dynamics of MBHs even in such high-redshift mergers.
Finally, we leave out a group of the most complex (with $>3$ MBHs in the remnant galaxy) merger systems for this work.
This group makes up a very small fraction of the total merger population (34 out of 2107).
In future works, it is still worthwhile to study these systems, as they often reside in the high-density peaks of the Universe and may trace the merger events during the formation of the first quasars.

In Figure \ref{fig:z6_pop_gal}, we show the host galaxy/halo information of the $z\sim 6$ mergers. 
The left two panels show the mass distribution of the dark matter, gas, and stellar components of the merger remnant at $z=6$.
The majority of $z=6$ mergers are between MBH seeds in dwarf galaxies, with a host halo mass of $10^{10}-10^{12}\,M_\odot$, and a host galaxy mass of $10^8-10^{10}\,M_\odot$.
At $z=6$, the merger host halos are often rich in gas, with the total gas fraction about ten times that of stars.
Previous works have shown that in such environments, the clumpy cold gas can result in the ejection of MBHs at kpc scales \citep[e.g.][]{Fiacconi2013, Tamburello2017} and result in early wandering MBHs.
Therefore, it is important to take gas physics into account when simulating mergers between the MBH seeds.

The right two panels of Figure \ref{fig:z6_pop_gal} show the power-law index of the density profiles and the densities measured at the \texttt{ASTRID} resolution limit \citep[similar to the method used in][]{Chen2022}.
Specifically, we assume that below a scale $r_{\rm ext}$ close to the resolution limit $2.8\times \epsilon_g=0.85\,{\rm kpc}$, the stellar density profile follows a single power-law $\rho \propto r^{-\gamma}$.
To measure the value of $\gamma$, we take the measured density from 10 bins just above $r_{\rm ext}$, and fit it to the power-law profile.
The gravitational potential of high-redshift galaxies is dominated by the dark matter halo above kpc scales.
In most cases, the dark matter density exceeds the stellar density by a factor of $\sim 10$.
However, these galaxies are gas-rich and mergers can also trigger a phase of rapid star formation.
Therefore, as we will also show later, the stellar densities are subjected to growth by a factor of $\sim 10$ over the timescale of a few hundred Myrs, and can dominate over dark matter on sub-kpc scales.

Finally, Figure \ref{fig:z6_pop_orb} shows eccentricities of the first orbit between the MBH pairs during the galaxy merger in \texttt{ASTRID}.
This is measured from the pericentric and apocentric separation between the MBH pair, and may be different from the Keplerian orbital parameters of the galaxy mergers.
The initial eccentricities have a wide distribution, with most ranging between 0.5 and 0.8.
We note that about $\sim 20\%$ of the pair has an initial eccentricity below $0.5$, and these pairs may experience much longer time (up to $\sim 2$ times longer than a pair with an initial eccentricity of $\sim 0.8$) in the dynamical friction phase before the formation of a hard binary \citep[e.g.][]{Gualandris2022}.
It is therefore important to include this population in the study of the seed sinking time.

\subsection{Resimulation System Selection}

As was described in the previous section, we categorize the merger systems according to the host environment complexity and use the number of MBHs in the merger remnants' host halo as a proxy for the complexity.
To obtain a good representation of different seed MBH merging environments, we sample merger events from all three categories for high-resolution idealized galaxy merger simulations.
We will resimulate a total of 15 \texttt{ASTRID} $z\sim 6$ mergers, including 7 in isolated galaxies, 6 in galaxies with new infalls at $z=6$, and 2 in multiple-galaxy interactions.
These systems are all chosen randomly from each population, to cover a statistical representation of all the merger events.

We show the sampled merger events on top of the overall $z\sim 6$ MBH merger population in Figures \ref{fig:z6_pop_bh}, \ref{fig:z6_pop_gal}, \ref{fig:z6_pop_orb} with the corresponding labels.
The colors represent the sub-population that each system belongs to (isolated, isolated with new infall, multiple galaxies).
For Figures \ref{fig:z6_pop_bh} and \ref{fig:z6_pop_orb}, the y-values are randomly chosen to spread out the data points for better visibility, while the x-values represent the MBH and orbital properties of the systems.
All the selected systems are mergers between two seed-mass MBHs with $M_{\rm BH} < 10^6\,M_\odot$.
They cover a wide range of galaxy, MBH, and orbital properties with high probability density in the parameter space.

In Table \ref{tab:astrid}, we list the detailed properties of the resimulation systems before and after the galaxy mergers.
For each system with an MBH merger at $z\sim 6$, we trace the host galaxies back to the snapshot before their interactions at $z_{\rm init}$ to initialize the resimulation IC.
For the selected systems, this corresponds to either $z=9$ or $z=7.6$.
During the galaxy merger, the total dark matter mass in the halo grows by a factor of $\sim 2$ for most systems due to further matter clustering.
The galaxy masses grow more significantly because of star formation: the galaxy remnant mass is usually an order of magnitude higher than the sum of the two parent galaxies.
We found that the SFR grew most rapidly during the galaxy merger.
Finally, all of the host galaxies are gas-dominated, with gas masses much larger than the stellar masses.

%% file: Sec4_ResimIC.tex
\section{Resimulation Set-up}
\label{sec:multigal}

\begin{figure*}
    \centering
    \includegraphics[width=0.75\textwidth]{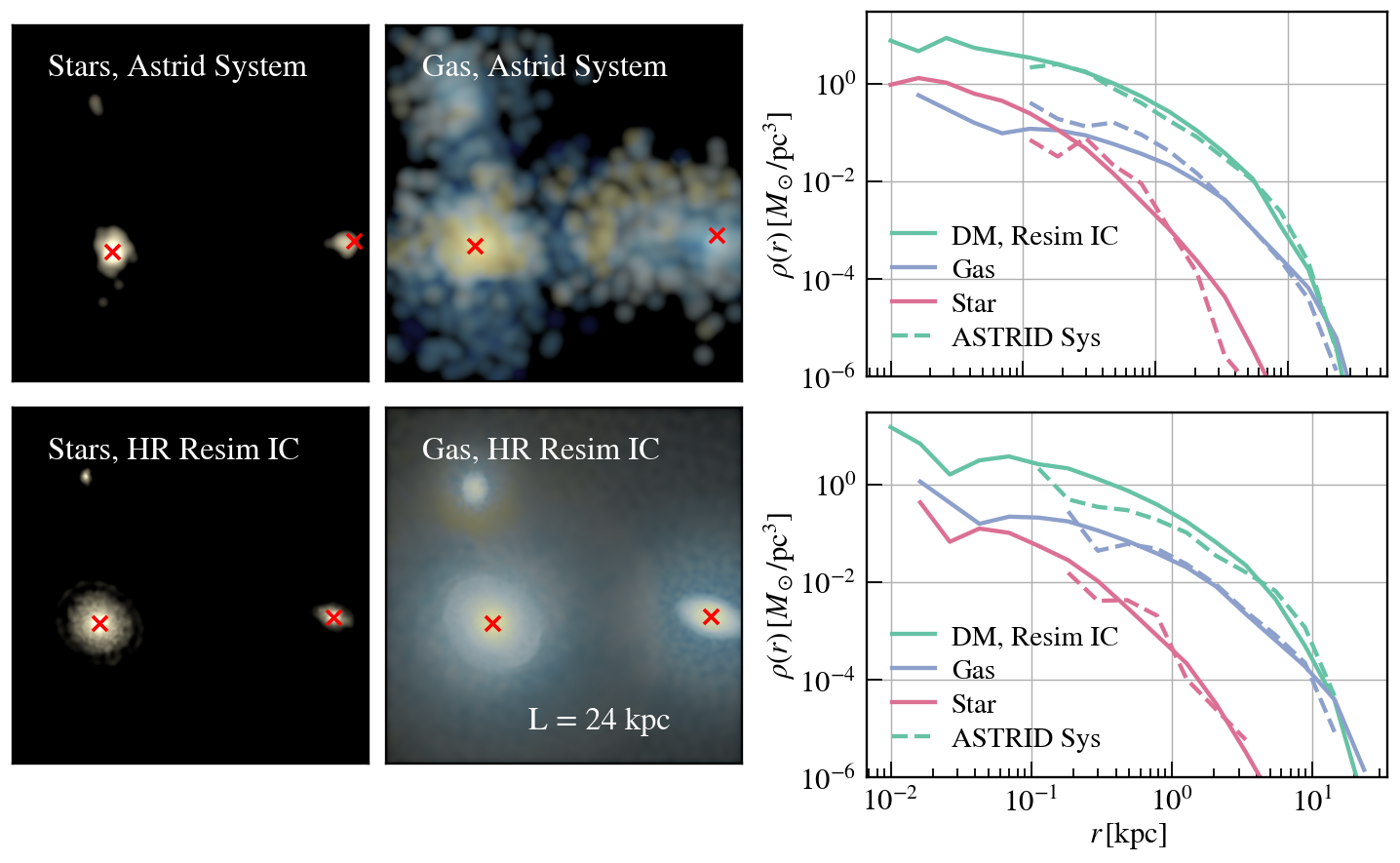}

    \caption{\textit{Left column:} visualization of stars in the \texttt{Astrid} merging galaxies and MBHs (\textbf{top}) compared with the IC of the high-resolution resimulation (\textbf{bottom}). The background brightness corresponds to the stellar density, with matched color scales between the top and bottom panels. Two merging MBHs are shown as \textbf{red crosses} on top of their host galaxies. \textit{Middle column:} Visualization of the gas environment in the \texttt{Astrid} system and the resimulation IC. The brightness represents gas density, and the colors represent the temperature (bluer colors are colder gas). \textit{Right column:} density profile comparisons between the \texttt{Astrid} galaxies (\textbf{dashed lines}) and the resimulation galaxies (\textbf{solid lines}). We compare the profiles of all three components (dark matter in \textbf{green}, gas in \textbf{blue}, and stars in \textbf{pink}), and show that the resimulation profiles matched well with the original profiles, but with extrapolations down to $>10$ times smaller scales than the original system.}
    \label{fig:resim_ic1}
\end{figure*}

\begin{figure*}
    \centering
        \includegraphics[width=0.90\textwidth]{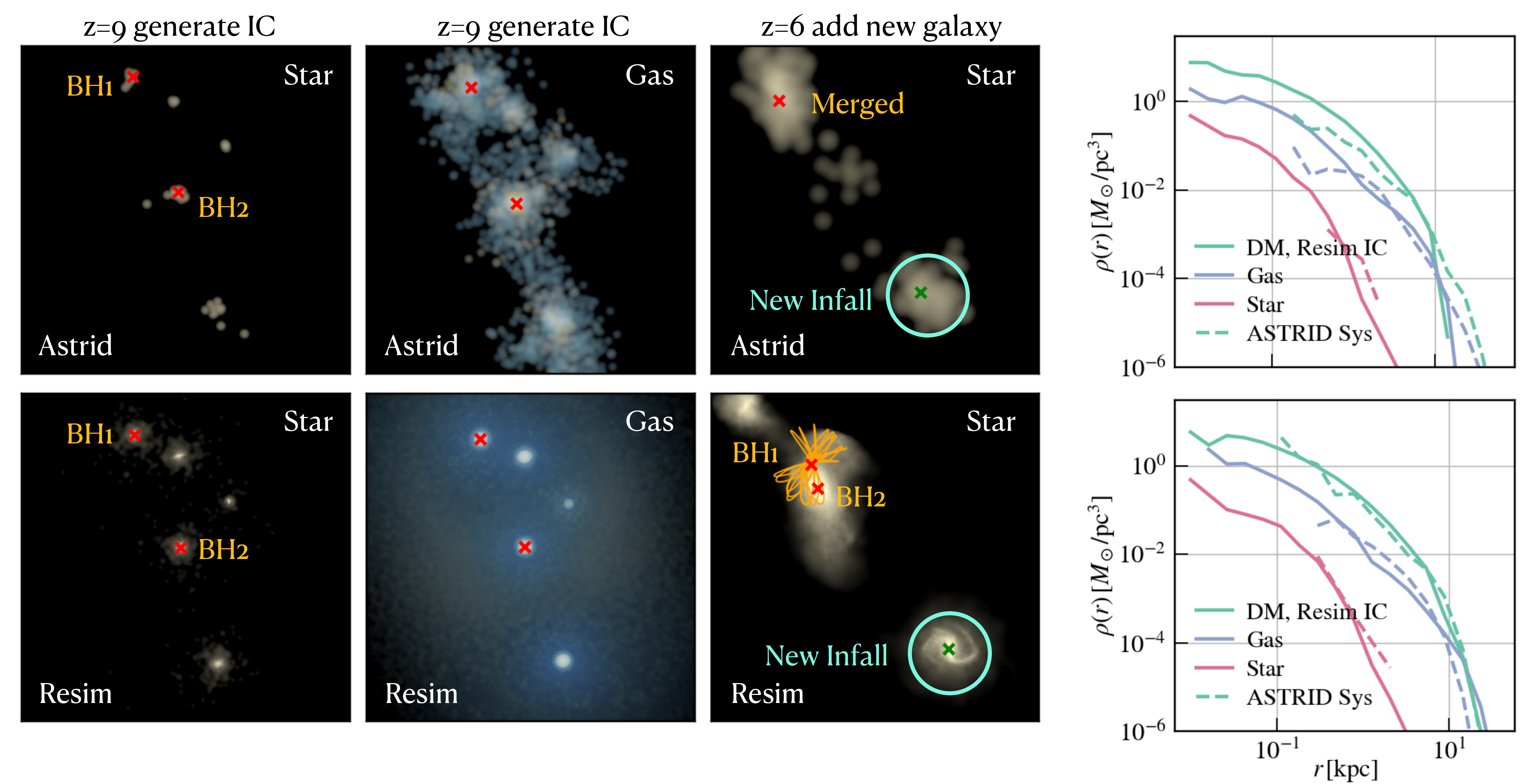}
    \caption{Similar to Figure \ref{fig:resim_ic1}, but for a more complex system with multiple galaxies in the IC as well as two new infalling MBHs and galaxies before the MBH pairs merge in the simulation. \textit{Right column}: The density profiles of the two infalling galaxies in \texttt{Astrid} and the resimulation.}
    \label{fig:resim_ic2}
\end{figure*}

\subsection{Initial Conditions}
\label{sec:multi_ic}

As was described in Section \ref{sec:sim}, in our idealized simulations, each subhalo consists of a dark matter halo component characterized by an NFW profile, a gaseous halo component with a beta profile, an exponential gaseous disk and stellar disk, and a stellar bulge following a Hernquist profile. 
To best mimic the original \texttt{ASTRID} systems, we initialize each idealized halo/galaxy in the resimulation IC according to the measured properties of the subhalos from a snapshot of the \texttt{ASTRID} simulation.
Here we describe how we set the parameters in idealized galaxy ICs.

We initialize an idealized galaxy for each subhalo identified by \texttt{Subfind} in the \texttt{ASTRID} merging systems with stellar mass $>10^6\,M_\odot$ and dark matter mass $>10^9\,M_\odot$.
We set $M_{\rm vir}$ as the total subhalo mass of the \texttt{ASTRID} subhalo. 
The dark matter halo is initialized with an NFW density profile, with the inner slope controlled by the concentration parameter $c$, and with the halo spin initialized to a constant value $0.033$.
We find that at the current \texttt{ASTRID} resolution and for the dwarf galaxies, we do not have enough information in the central region to provide a good fit for $c$.
Thus we set $c=4$ to fit with the dark matter density profile at the high-redshift regime of this work \citep[see e.g.][]{Prada2012} and find that this value fits the profiles well on the \texttt{Astrid}-resolved scales. 
We note that the sinking time of seed MBHs can be sensitive to the inner DM density profiles \citep[e.g.][]{Tamfal2018}, and can potentially be used to distinguish between different dark matter models.
However, this should not affect our major conclusions as we are only sampling from a single cosmology.

In \texttt{ASTRID}, there is no explicit gas disk (especially at high redshifts), and so we set the mass of disk gas according to the fraction of star-forming gas in the \texttt{ASTRID} subhalo.
The rest of the gas is put into the gaseous halo component.
We assume exponential, rotation-supported disk gas with scale lengths fitted to the original system's density profile, and we fix the scale height at 0.2 times the scale length.
The gas temperatures are initialized to pressure equilibrium \cite{Springel2005b}.
We also initialize a hydrostatic gas halo according to a beta profile with $R_c/R_s = 0.5$ and $\beta = 0.4$ (we tested that the dynamics of MBHs are not very sensitive to this choice, and defer the detailed study of its effect to future works). 

The stellar disk and bulge fraction are decomposed following the kinematic decomposition algorithm in \citep[e.g.][]{Abadi2003, Scannapieco2009}.
The stellar disk follows the same profile as the gas disk, and the stellar bulge follows the Herquist profile with scale length set according to the half-mass radius of the \texttt{ASTRID} galaxy: $a = r_{\rm half}/(1+\sqrt{2})$ \citep{Hernquist1990}.
We relax the initial conditions for each galaxy adiabatically and in isolation for $200\,{\rm Myrs}$, before assembling the merging system and putting in the MBHs.
After relaxation, we assemble all galaxies in the system according to their relative positions, velocities, and the direction of the rotation vectors originally found in \texttt{ASTRID}.
We also add MBH particles according to their positions, velocities, and masses in \texttt{ASTRID}.

Figure \ref{fig:resim_ic1} shows the comparisons between the \texttt{ASTRID} merging system and the resimulation initial conditions generated as described above.
In the left and middle columns, we show visualizations of the host galaxies and gas environments, with matched color scales between \texttt{Astrid} and the resimulation (brightness corresponds to the density, and gas is color-coded by temperature with blue corresponding to cold gas). 
The resimulation IC resembles the morphologies, stellar densities, and gas temperature of the original system, but shows significant improvement in mass and spatial resolution.
The improvement in resolution can be seen more clearly from the left column, where we show the density profiles of the original halos and the resimulation halos (measured after the adiabatic relaxation, just before we start the resimulations).
The detailed properties of all components in each recovered galaxy are shown in Appendix \ref{app:ic}.

\subsection{Other In-falling Galaxies/BHs During the MBH Pairing}
\label{sec:multi_infall}

When we resimulate the cosmological merging MBHs at a much higher resolution, the resolved DF timescale between galaxy mergers and MBH mergers may lengthen, both because we do not boost the dynamical mass of the BH particles and because we use a stricter merging criterion with smaller softening lengths.
One result of the longer DF timescale is the infall of other galaxies and BHs to the merging system that may either intervene or accelerate the orbital decay of the original pair.
To fully mimic the cosmological system, we need to take these newcomers into account, as they can both interfere with the original mergers and alter the properties of the remnant host galaxies.

To treat this scenario, we put new galaxies into the resimulated system at the time ($t_1$) when we see another galaxy with a BH coming within two times the virial radius of the original system.
Similar to how we set up the original system's initial condition, we first initialize the new galaxies based on the properties of their cosmological counterpart at $t_1$.
Then we compute the position ($\Delta \mathbf{x}(t_1)$) and orientation ($\Delta \theta(t_1)$) of the new galaxy relative to the original galaxy. 
 To keep the total momentum of the new system at zero, we compute the velocities of both the original galaxy ($\Delta \mathbf{v_{\rm old}}(t_1)$) and the new galaxy ($\Delta \mathbf{v_{\rm new}}(t_1)$) with respect to the COM of the combined system. 
Finally we add the new galaxy at ($\Delta \mathbf{x}(t_1)$, $\Delta \mathbf{v_{\rm new}}(t_1)$, $\Delta \theta(t_1)$) to the resimulation, and modify the total velocity of the original galaxy to be $\Delta \mathbf{v_{\rm old}}(t_1)$.

\begin{figure*}
    \centering
    \includegraphics[width=0.99\textwidth]{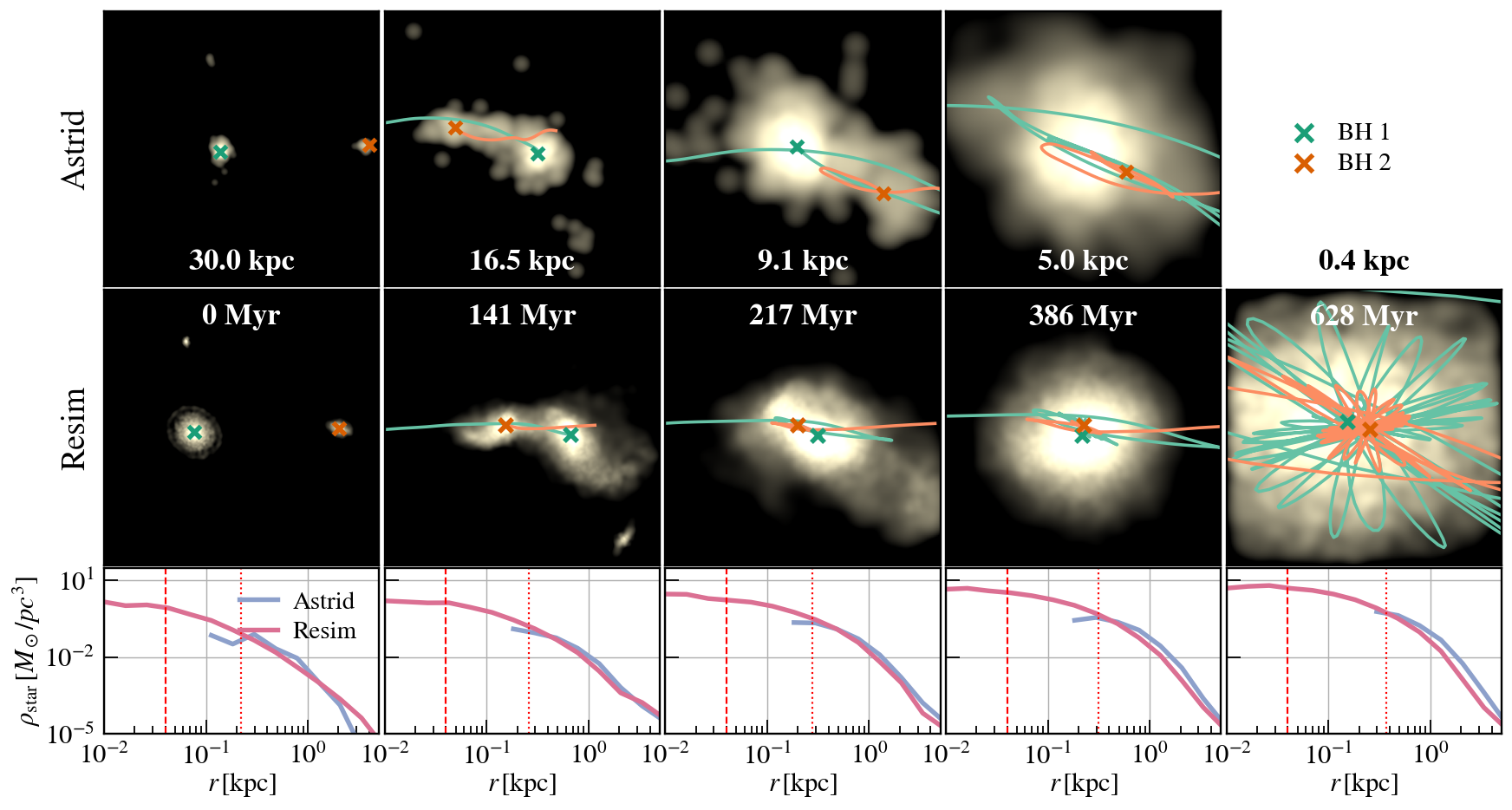}
    \caption{\textit{Top row:} trajectories of the MBH pair (crosses) plotted on top of the merging host galaxies in \texttt{ASTRID} \texttt{system3}. The simulation merger happens between the third and fourth frames.
    \textit{Middle row:} evolution of the same system in the high-resolution resimulation. The large-scale galaxy merger and MBH orbital properties are paralleled, but the orbits are resolved down to $\sim 20\,{\rm pc}$ scales, close to the binary hardening (last frame). \textit{Bottom row:} comparison between the stellar density profiles of the primary galaxy (first two frames) and the remnant galaxy (last three frames) in \texttt{ASTRID} (\textbf{purple}) and the resimulation (\textbf{pink}). The density profiles in the resimulation match well with the \texttt{ASTRID} system, with an extrapolation to $>10$ times smaller scales.}
    \label{fig:gal_evol1}
\end{figure*}

Figure \ref{fig:resim_ic2} shows an example system where we add new infalling galaxies and BHs during the resimulation, based on the information from the \texttt{ASTRID} system.
This system originally consists of a merger between five dwarf galaxies with two seed MBHs.
After we evolve the resimulation for $\sim 300\,{\rm Myrs}$ down to $z=6$, we observe a new infalling galaxy that is about to merge with the original system in \texttt{ASTRID} (the green cross and the galaxy associated with it in the top row, third column of Figure \ref{fig:resim_ic2}).
We initialize this galaxy following the procedures described above and add it into the resimulation (bottom row, third column), so that it will start to interact with the original MBH pair.
The right columns show the density profiles of the new galaxies in the simulation compared with \texttt{ASTRID}, and again we can see that we match the \texttt{ASTRID} galaxy/halo profiles well on the $>{\rm kpc}$ scale, while achieving more than ten times better spatial resolution.

We note that during the resimulation, the mass of the original system does not grow, and hence the new galaxies may fall into different potentials in the cosmological simulation and the resimulation. 
In general, we verify that the total mass of the original system does not grow by more than a factor of three before the injection of new galaxies.
We defer more detailed investigations of this effect and careful treatments of the mass growth to future works.

%% file: Sec5_Result.tex
\section{Results}
\label{sec:result}

Using the method described in Section \ref{sec:multigal}, we set up a total of 15 galaxies and MBH merger initial conditions for the chosen \texttt{ASTRID} merging systems shown in Section \ref{sec:astrid_pop}, and with the ``full-physics" subgrid physics models depicted in Section \ref{sec:sim}.
In this section, we show the results from these high-resolution resimulations.

.
\subsection{Evolution of the host galaxies}

Although the initial conditions for the resimulations are set to match the \texttt{ASTRID} system as closely as possible, it is not guaranteed that their subsequent evolution will be similar.
As a first test, we want to make sure that the general properties of the galaxies and MBH orbits in the resimulations still mimic the evolution in \texttt{ASTRID} to at least the \texttt{ASTRID} MBH merger time.
Only in this case can we draw further comparisons and connections between the cosmological simulation and idealized galaxy merger simulations.

In Figure \ref{fig:gal_evol1}, we show a parallel comparison between the galaxy merger and BH orbits in \texttt{ASTRID} and in the resimulation system for an isolated galaxy merger (\texttt{system1}).
On large scales, we find a good match between the progress of the galaxy merger and MBH orbits between the two systems (also shown later in Figure \ref{fig:orbits_all}).
The \texttt{ASTRID} system merged in the fourth frame, while in the resimulation we further evolve the orbits down to $\sim 20\,{\rm pc}$, and the sinking timescale is much longer (fifth frame).
The bottom panels show the evolution of the stellar density profiles in both systems.
The overall density profile evolution of the resimulation system matches well with the \texttt{ASTRID} system over $\sim 1\,{\rm Gyr}$, because we also try to match the gas properties in the resimulation initial conditions.
More importantly, we note that the central density grows by a factor of $\sim 10$ during the orbital decay time of the MBH pair.
The growth in central stellar density can significantly impact the dynamics of the BHs on sub-kpc scales.

\begin{figure*}
    \centering
    \includegraphics[width=0.99\textwidth]{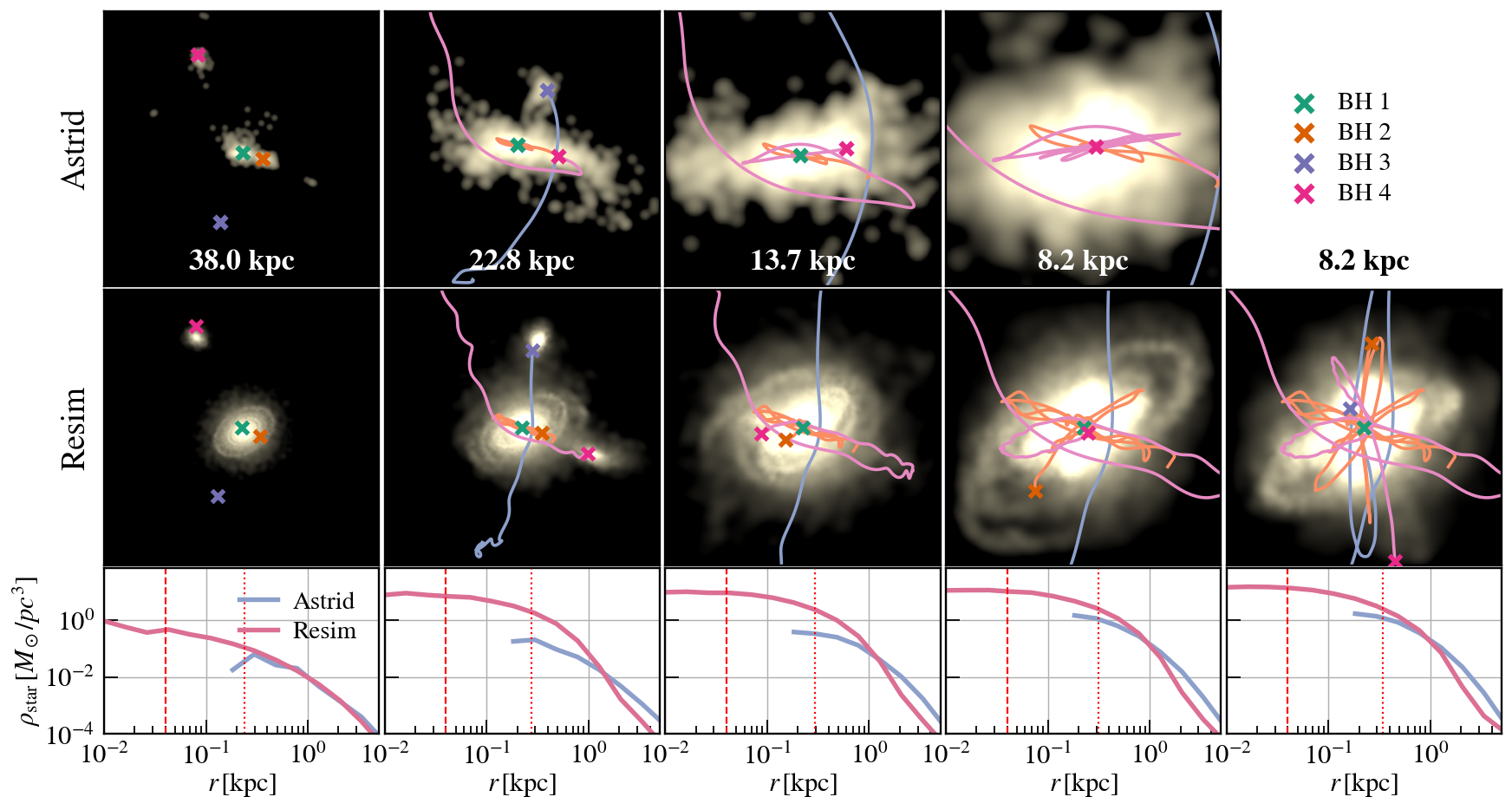}
    \caption{Similar to Figure \ref{fig:gal_evol1} but for a system with multiple galaxy mergers (\texttt{system15}). The MBH orbits are more stochastic for this system, and the orbit of BH2 (orange) widens with the infall of BH3/BH4 and their host galaxies.}
    \label{fig:gal_evol2}
\end{figure*}

Figure \ref{fig:gal_evol2} shows the galaxy merger comparison for a more complex system with multiple galaxies merging (\texttt{system15}).
In this case, the \texttt{ASTRID} system goes through two consecutive MBH mergers within $\sim 350\,{\rm Myrs}$(between BH1 with BH2, then with BH4).
In the resimulation system, the MBHs have a difficult time merging: BH3 and BH4 (and their host galaxies) will start to interfere with the orbits of BH1 and BH2 before they can merge.
As a result, BH2 is disrupted to a wider orbit (column 4) where the dynamical friction becomes inefficient.

\begin{figure}
    \centering
    \includegraphics[width=0.48\textwidth]{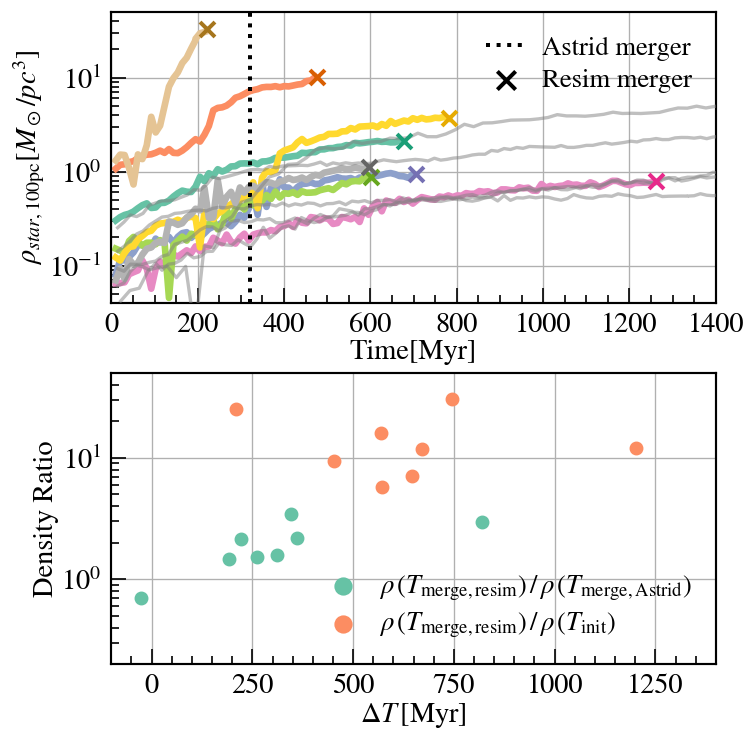}
        \caption{\textit{Top panel:} Evolution of the central stellar density (measured at $100\,{\rm pc}$ from the most massive galaxy center) during the MBH inspiral and merger in the simulations.The colored lines show the density evolution until the resimulation merger time (\textbf{crosses}) for systems that merged in the resimulations. The \textbf{thin grey lines} are systems that stall in the resimulations. The vertical dotted line marks the \texttt{ASTRID} merger time. \textit{Bottom panel:} the ratio between stellar density measured at the resimulation merger and the initial condition (\textbf{green}), and between the resimulation merger and the \texttt{ASTRID} merger (\textbf{orange}). }
    \label{fig:sfr_evol}
\end{figure}

We summarize the evolution in central stellar density across all resimulated systems in Figure \ref{fig:sfr_evol}, from the start of the resimulation ($\sim 300\,{\rm Myrs}$ before the \texttt{ASTRID} merger) to the resimulation merger time.
The thick colored lines show the systems that merged in the resimulation, while the thin grey lines show the systems that stall at $\sim kpc$ scales for more than $1.5\,{\rm Gyr}$.
For almost all systems the stellar densities increase by an order of magnitude during the MBH sinking.
After the \texttt{ASTRID} merger, the stellar density still increases by a factor of $2\sim 3$.
This direct comparison implies that one should account for the newly formed stars when using post-processed analytical models to compute the binary hardening time for mergers in cosmological simulations, and the resulting hardening efficiency may increase.
We also note that for the system with the steepest increase in central stellar density (\texttt{system11}, shown in light brown), the MBH sinks even more efficiently in the resimulation system and merges before the \texttt{ASTRID} merger takes place.

The recent work by \cite{Liao2023b} has shown that the increase in central stellar density and the development of a nuclear stellar core can significantly increase the binary hardening efficiency in SMBHs.
Our result again highlights the effect of central star formation on MBH sinking timescales, in the context of high-redshift seed-mass MBH mergers.
In the resimulation runs, we include the ``full-physics" modeling of both star formation and AGN feedbacks (thermal and kinetic), and thus the stellar density profile flattens (similar to the ``CoolStarKinAGN" modeling in \cite{Liao2023b}).
Nonetheless, we still find very high star formation rates (up to $50\, M_\odot/{\rm yr}$) and the development of stellar nuclei when the merging galaxies are compact enough (e.g. \texttt{system11}), and the MBH seeds can sink very efficiently in these cases.

%-----------------------------------------------------------------------------------------
\subsection{Seed MBH merging timescale}

\begin{figure*}
    \centering
    \includegraphics[width=0.98\textwidth]{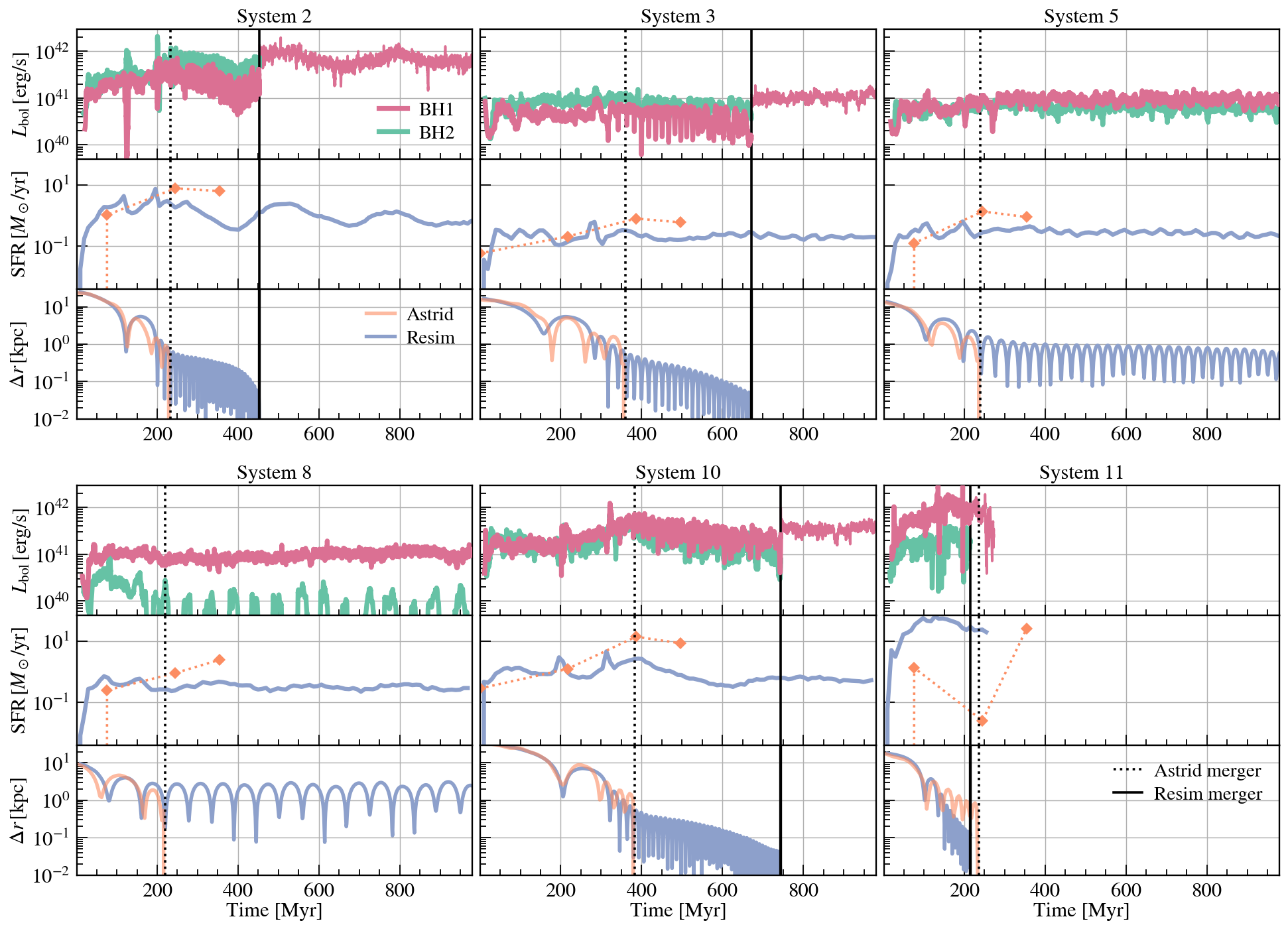}
    \caption{Evolution of the AGN luminosities (top row in each panel), star formation rate (middle row in each panel), and MBH pair separation (bottom row in each panel) in the resimulation.
    For the SFR and orbital separation we also compare the resimulation (\textbf{purple}) with the original \texttt{ASTRID} systems (\textbf{orange}).
    We show six systems representative of the orbital properties of the fifteen resimulations. \texttt{system2}, \texttt{system3}, \texttt{system10}, and \texttt{system11} go through efficient orbital decay, while  \texttt{system5}, \texttt{system8} stall at the kpc scale. The SFRs in the resimulations resemble those of the \texttt{ASTRID} system well during and after the \texttt{ASTRID} merger. The initial few orbits also show a good resemblance to the \texttt{ASTRID} orbits.}
    \label{fig:orbits_all}
\end{figure*}

In \texttt{ASTRID}, MBH orbits are resolved down to scales of $\sim 1\,{\rm kpc}$, and so MBH pairs are assumed to ``merge" after that.
However, in high-redshift dwarf galaxies, it is typical for seed MBHs to stall on kpc scale orbits for over a few Gyrs \citep[e.g.][]{Ma2021, Pfister2019, Partmann2023}.
In this section, we study the merging timescales and the stalling of MBH seeds of the systems in the resimulation, and the correlation with large-scale orbital and galaxy properties.

Figure \ref{fig:orbits_all} shows the AGN luminosities, star-formation rate, and MBH pair separation of six resimulation systems.
These systems are chosen to cover the range of galaxy and the orbit properties across the 15 resimulated systems.
Out of the six systems, four go through relatively fast orbital decay and merge within the resimulation after $<800\,{\rm Myrs}$, while two systems show stalling at kpc separations for over a Gyr.
Notably, the initial few orbits of the resimulation agree well with the \texttt{ASTRID} pair, even though we alleviate the boost in dynamical mass on the seeds (as we will show later, this is because the first few orbits are governed by the gravitational potential).
This agreement indicates that cosmological simulations with well-calibrated dynamical friction treatment faithfully model the initial orbital properties of the MBH pairing.
Such orbital properties can provide useful initial conditions for subsequent orbital evolution or analytical modeling of the MBH merging timescales \citep{Gualandris2022}.

The middle panels in Figure \ref{fig:orbits_all} compare the resimulation SFR with that of the corresponding \texttt{ASTRID} systems.
We find general agreement between the two before and after the \texttt{ASTRID} merger.
% We did not follow the \texttt{ASTRID} SFR further, because the MBH merger can result in an increase in AGN activity and start to affect the host galaxy SFR.
In almost all systems, we find an increase in AGN activity and star formation rate associated with the first few pericentric passages.
In particular, fast orbital decays are associated with stronger AGN activities (\texttt{system2}, \texttt{system10}, \texttt{system11}).
This comes as no surprise since these systems are also on the high-mass, high-density end of the galaxy population (see e.g. Figure \ref{fig:z6_pop_gal}).

We note that in the resimulation we still do not resolve the full dynamical range until the MBH coalescence, and so the MBHs ``merge" when their orbital separation is $\sim 20\,{\rm pc}$.
To validate that the MBHs' motion is not affected by numerical noise above the merging distance, we measure the wandering radius of the merger remnant following \cite{Bortolas2016}, by averaging the mean displacement of the merger remnant from the galaxy center over time. We find that for all systems that merged, the remnant MBH has a mean displacement of $\sim 20\,{\rm pc}$.
By numerically merging the MBH at this separation, we pick out systems that will likely form a bound binary, since the stalling in the dynamical friction regime is seen at $\sim {\rm kpc}$ scales \citep[e.g.][]{Gualandris2022, Partmann2023, Koehn2023}.
Further stalling may happen at $\sim pc$ scales due to the depletion of the loss cone, but here we only focus on the pair evolution in the dynamical friction regime and defer the smaller scale dynamics to future works.

\begin{figure*}
    \centering
    \includegraphics[width=0.43\textwidth]{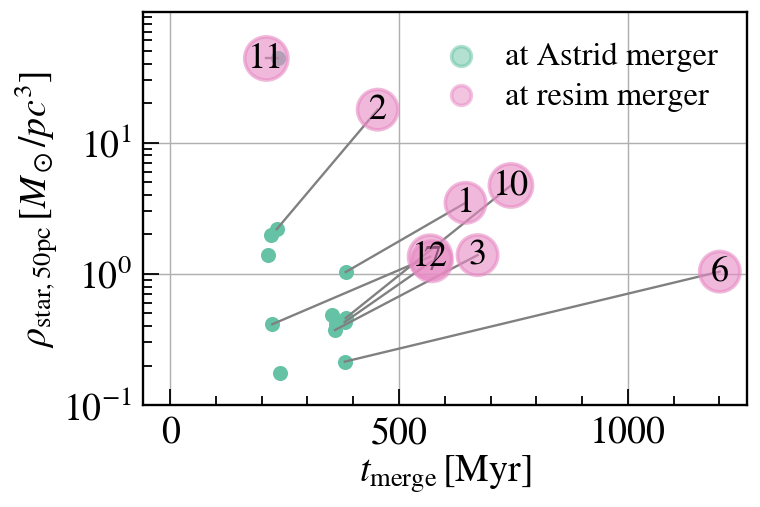}
    \includegraphics[width=0.43\textwidth]{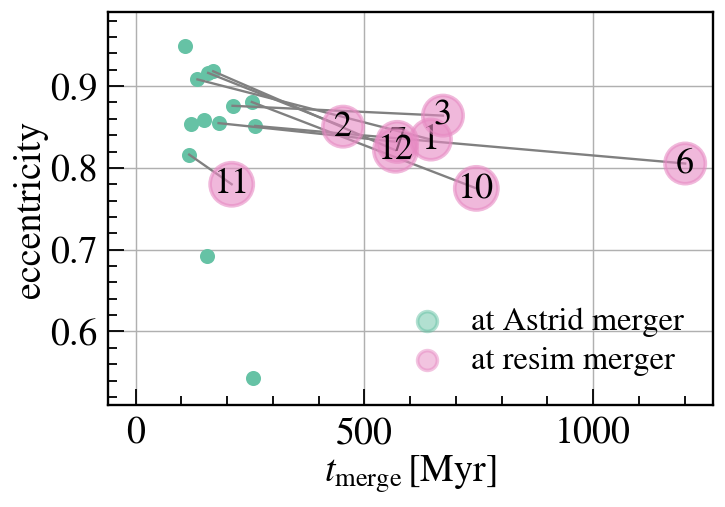}
        \caption{\textit{Left panel:} relation between MBH merging timescale and stellar density at $50\,{\rm pc}$ from the galaxy center. We measure density both at the \texttt{ASTRID} merger time (\textbf{green}) and at the resimulation merger time (\textbf{pink}). We see a tight correlation between the merging timescale and stellar densities. We plot densities in systems that do not merge within 1.5 Gyrs in the resimulation on the right of the box. \textit{Right panel:} initial (\textbf{green}) and final (\textbf{pink}) orbital eccentricity of MBH pairs in the resimulation and the correlation with the merging timescale. The eccentricity clusters around 0.8 when the pair starts entering into the hardening phase. }
    \label{fig:tmerge_1d}
\end{figure*}

The left panel of Figure \ref{fig:tmerge_1d} summarizes the correlation between galaxy properties and seed MBH merging timescales in all resimulation systems.
Each system is labeled with the corresponding number in the plot.
Out of 13 resimulated systems in isolation, 8
MBH pairs will merge within 1.5 Gyrs (i.e. before $z\sim 3$).
We find a strong correlation between the merging timescales and the central stellar density, both at the \texttt{ASTRID} merging time and the resimulation merging time.
We note that in our simulations, we do not hold any other galaxy or MBH properties constant while varying the stellar densities.
Therefore, this correlation is a result of marginalizing over other parameters in the merger \citep[see also e.g.][]{Tremmel2018}.
In this plot, we leave out the two systems with multiple MBHs (\texttt{system14} and \texttt{system15}), as the MBH dynamics are more complex in those cases and a simple scaling with stellar density may not apply. We will discuss these systems in a later section.

Besides the stellar density, it is known that the orbital eccentricity also has a large impact on the MBH merging timescales, both in the dynamical friction regime \citep[e.g.][]{Taffoni2003, Gualandris2022} and in the loss-cone scattering regime \citep[e.g.][]{Sesana2010}.
In the right panel of Figure \ref{fig:tmerge_1d}, we show the eccentricity evolution between the \texttt{ASTRID} MBH merger time and the resimulation merger time.
For the systems that merged efficiently, the orbital eccentricity from at \texttt{ASTRID} merger falls above $0.8$, and we find slight circularization during the subsequent dynamical friction phase.
At the resimulation merger time, the eccentricities of the MBHs all fall close to a value of $\sim 0.8$.
The two systems with $e<0.8$ at the \texttt{ASTRID} merger time do not merge in the resimulation within $1.5\,{\rm Gyr}$ (\texttt{system4} and \texttt{system9}).
These results imply that the high-redshift seed mergers more likely come from MBH pairs with high initial orbital eccentricities, and would retain these high $e$ values by the end of the dynamical friction phase.

\begin{figure}
    \centering
    \includegraphics[width=0.49\textwidth]{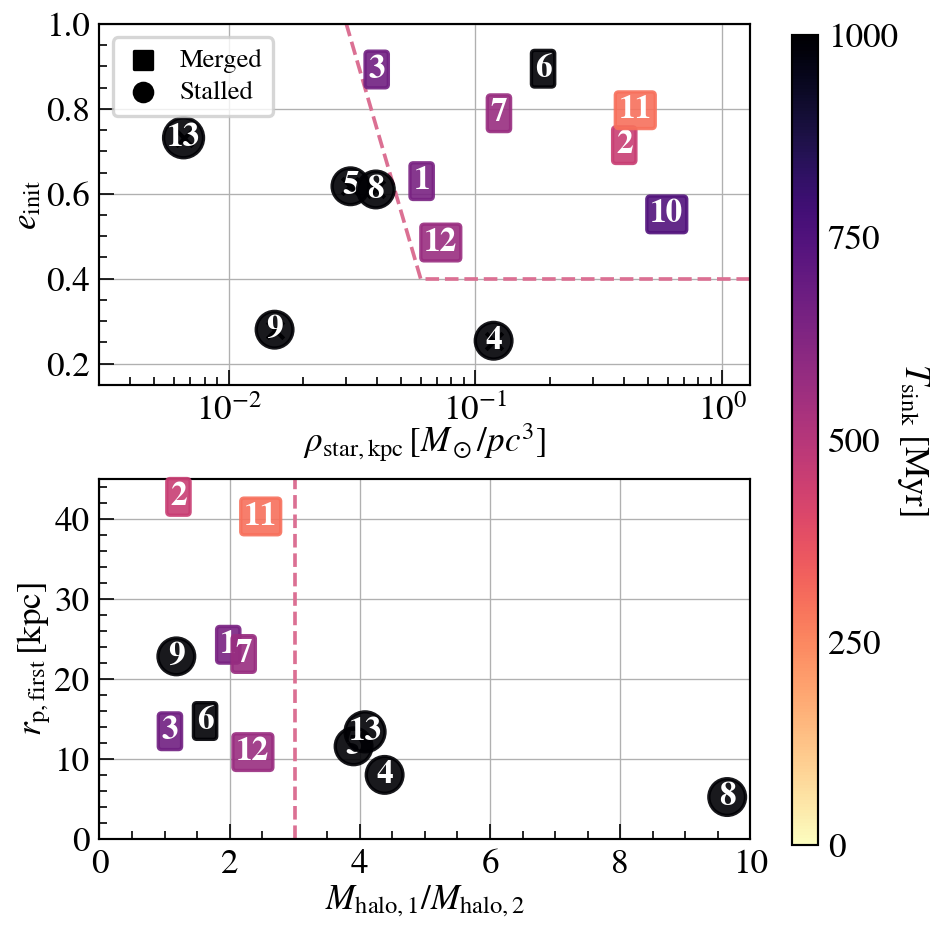}
        \caption{Sinking time of MBH seed pairs for the thirteen mergers simulated in isolation. \textit{Top panel}: sinking time on the plane of stellar density and initial eccentricity of the MBH pair in \texttt{ASTRID}.  \textit{Bottom panel}: sinking time on the plane of halo mass ratio and pericentric radius between the galaxies (computed based on relative velocities and positions). In eight systems (\textbf{squares}) the MBH merges in the resimulation in $\sim 1.2 Gyr$ (i.e. by $z\sim 3.5$). The colors indicate the sinking time of each system that merged. Five pairs do not merge in the simulation (\textbf{black circles}). The merged systems are mostly characterized by high stellar density, high orbital eccentricity, and major halo mergers.}
    \label{fig:tmerge_2d}
\end{figure}

\subsection{MBH mergers and large-scale properties}
\begin{figure}
    \centering
    \includegraphics[width=0.49\textwidth]{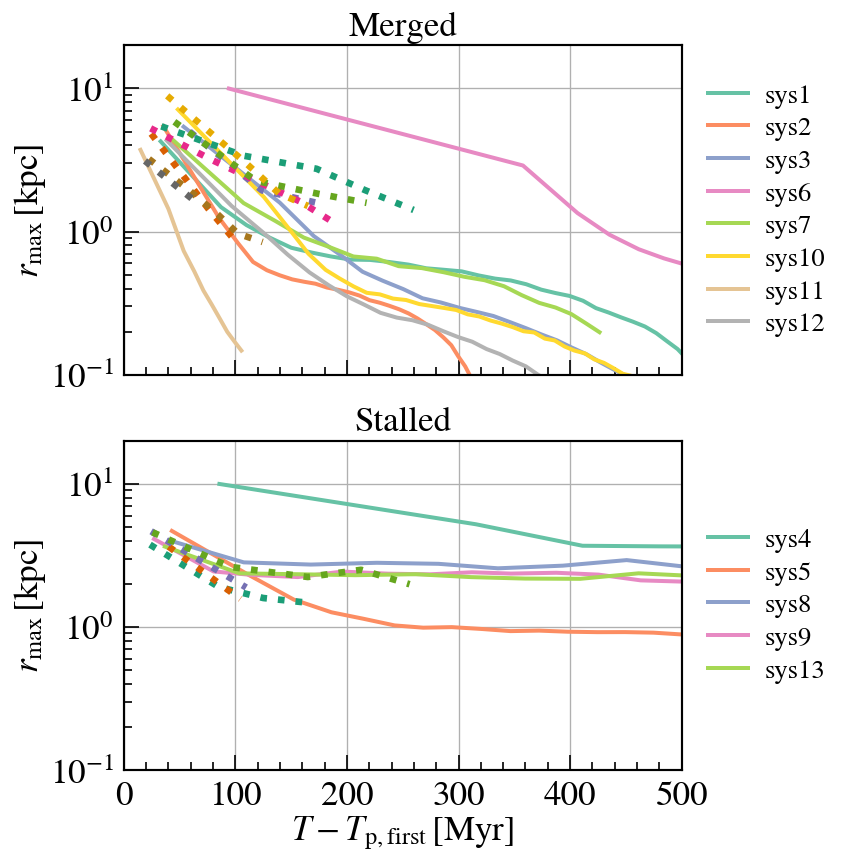}
        \caption{Evolution of apocentric distances of the secondary MBH $r_{\rm max}$ since the first pericentric passage in systems that merged in the simulation (\textit{top}) and systems that stalled (\textit{bottom}). We show the comparison between the distances in the original \texttt{ASTRID} systems (\textbf{dotted lines}) and the resimulated systems (\textbf{solid lines}).}
    \label{fig:rmax}
\end{figure}

One main motivation for using cosmological simulations to set up resimulations of galaxy and MBH mergers is to understand whether we can use the information from cosmological simulations to predict the dynamics of MBHs at sub-resolution scales.
In this section, we connect each resimulation and the MBH merging time back to properties of the ASTRID-resolved quantities and investigate what would be a good indicator for the sub-resolution dynamical behavior of MBH seeds.

Figure \ref{fig:tmerge_2d} shows the quantities from the \texttt{ASTRID} system that we found most correlated with the orbital decay timescale in the resimulation.
From the top panel, we see that systems that merge in the resimulation are characterized by high initial eccentricity between the MBH pair ($\gtrsim 0.4$), and high stellar density at $\sim {\rm kpc}$ scales ($\gtrsim 5\times 10^{-2}\,M_\odot/pc^3$).
The bottom panel shows the relation between the sinking timescale and the properties of the host halo mergers.
$r_{\rm \, first}$ is the pericentric radius of the initial galaxy merger, computed from the halo masses, initial relative positions and velocities of the two galaxies.
We note that in some cases there are more than two halos involved in the merger, and so the orbits cannot be exactly characterized by a Keplerian orbit.
The stalled seeds are mostly found in minor halo mergers with small pericentric radii.
In these mergers, the host halo and galaxy of the secondary MBH are most quickly disrupted, leaving the MBH completely bare from the very early stages.

We further investigate if the resolved first orbits in \texttt{ASTRID} show an indication of the subsequent orbital properties.
In Figure \ref{fig:rmax}, we plot the evolution of the apocentric radius $r_{\rm max}$ in all resimulated systems, since the first pericentric passage between the two galaxies.
We separate the systems by whether the MBH pair merged or stalled in the resimulation.
A comparison between the top and bottom panels shows that the merged systems typically started with lower orbital energies at the beginning of MBH pairing, and sink to $r_{\rm max}<1\,{\rm kpc}$ within $\sim 200\,{\rm Myr}$ of the first pericentric passage.
For systems that stalled in the resimulation, none of the MBH sink to $r_{\rm max}<1\,{\rm kpc}$ within the first $\sim 300\,{\rm Myr}$.
If the MBHs' initial orbital sizes are larger, they would experience less efficient dynamical-friction-driven decay due to the lower local densities, and as a result will stall on $r_{\rm max}\sim 1\,{\rm kpc}$ for longer than a Gyr.

\begin{figure}
    \centering
    \includegraphics[width=0.45\textwidth]{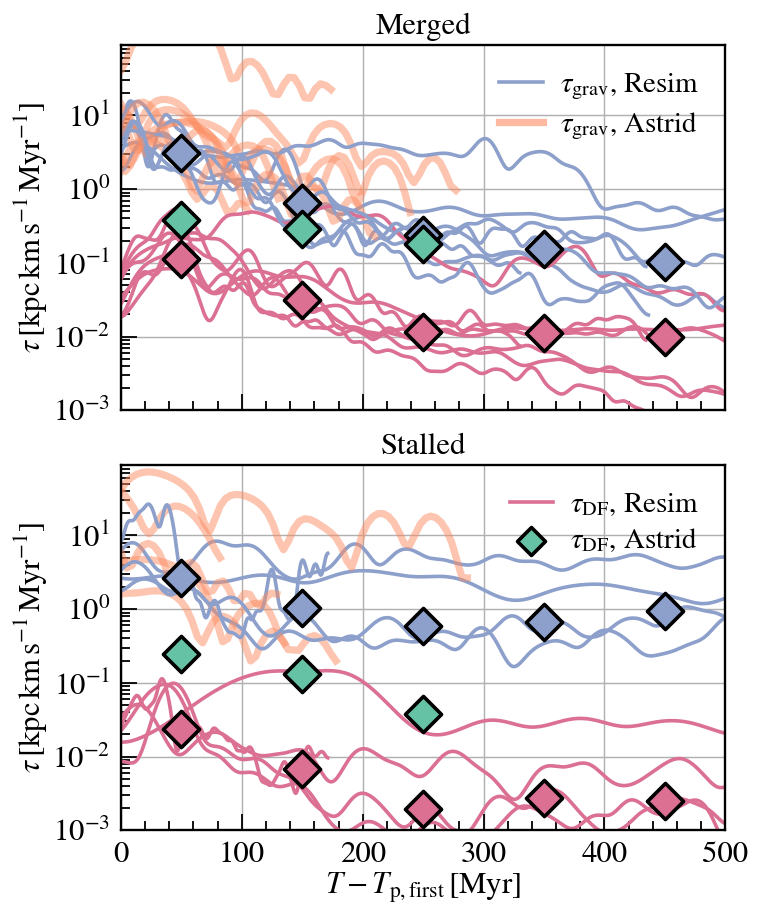}
        \caption{Time evolution of the gravitational torque on the secondary MBH since the first pericentric passage in \texttt{ASTRID} (\textbf{thick orange}) and the resimulation (\textbf{thin blue}). The \textbf{pink lines} show the dynamical friction torque from the resimulations. The gravitational torque in \texttt{ASTRID} is recovered by the resimulation in most systems, and it is two orders of magnitude larger than the dynamical friction torque, as was also shown in \citet{Bortolas2020}. The \textit{diamonds} show the median torque among each group within time bins of $100\,{\rm Myrs}$. The \textbf{green diamonds} are the median DF torque from \texttt{ASTRID}.}
    \label{fig:torque}
\end{figure}

\begin{figure*}
    \centering
    \includegraphics[width=0.98\textwidth]{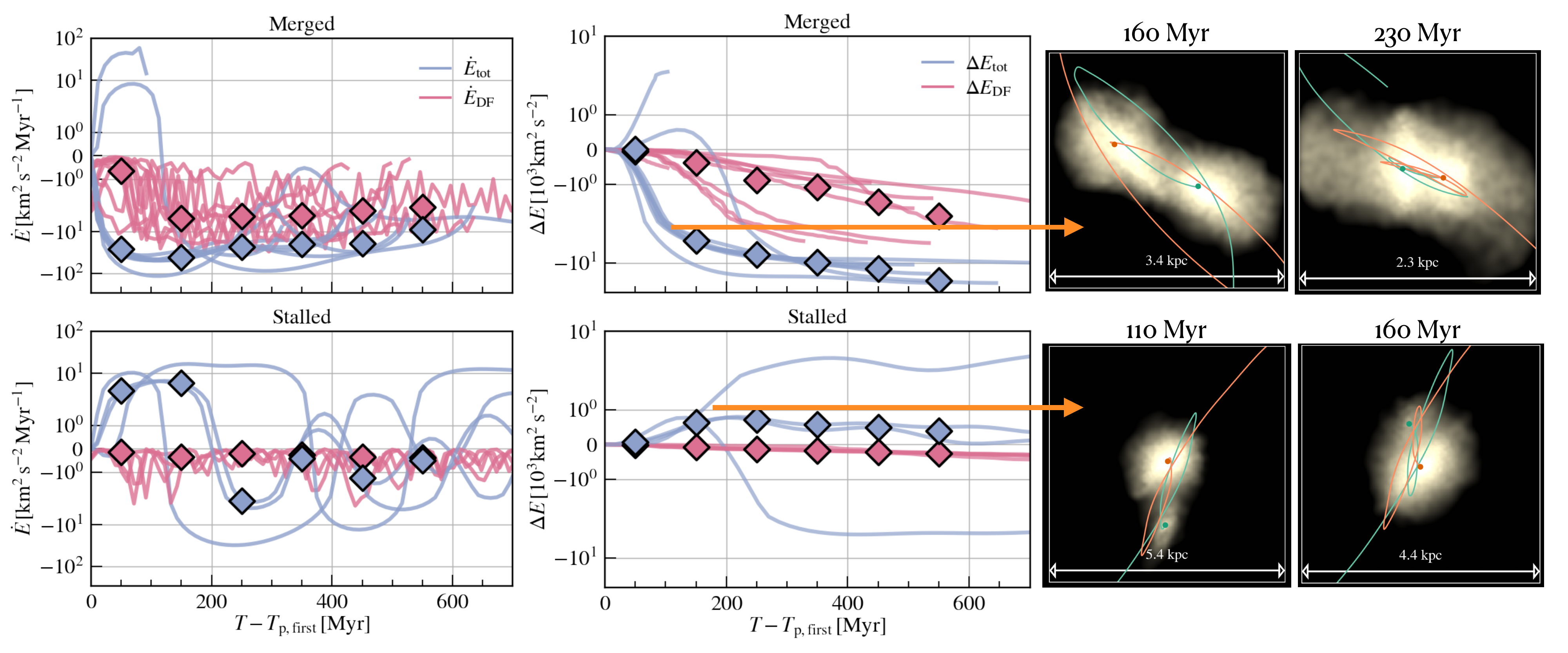}

    \caption{\textit{Left column:} Rate of total energy change of the secondary MBH (\textbf{blue}) compared with the energy loss rate due to dynamic friction (\textbf{pink}). The \textbf{lines} show each system and the \textbf{diamonds} are the median across all systems in each time bin. We plot the systems that merged in the resimulation in the \textit{top panel} and the systems that stalled in the \textit{bottom panel}. The merged MBHs experience loss of energy dominated by gravity, while the stalled MBHs gain energy during the first $\sim 200\,{\rm Myrs}$ of the galaxy merger. \textit{Middle column:} Cumulative change in the MBH energy since the first pericentric passage between the pair. \textit{Right column:} visualization of two galaxy mergers that lead to a merged pair (\textit{top}) and a stalled pair (\textit{bottom}). The stalled MBHs are mostly found in head-on collisions of minor galaxy mergers, in which the secondary host is quickly dissolved.}
    \label{fig:energy}
\end{figure*}

Motivated by the study in \cite{Bortolas2020}, we calculate the torque onto the MBHs at different times of the pairing, and from large-scale gravitational force to the local dynamical friction force.
In Figure \ref{fig:torque}, we show the magnitude of the total gravitational torque on the sinking MBHs from all resimulations compared with the dynamical friction torque.
The torque is calculated as a cross product between $\textbf{r}$ (distance to the primary galaxy center or the remnant galaxy) and $\textbf{F}_{\rm grav}$ or $\textbf{F}_{\rm DF}$.
$\textbf{F}_{\rm grav}$ is taken as the total resolved gravitational force on the MBH in the simulation, and $\textbf{F}_{\rm grav}$ is the subgrid-dynamical friction force computed at each MBH time step.
Corroborating the results shown in \cite{Bortolas2020}, we also find that the large-scale gravitational torque dominates the local dynamical friction force by $\sim 2$ orders of magnitude.
This is true both during the galaxy merger and during the subsequent sinking of the MBHs.
By splitting again between merged and stalled resimulation systems, we find that during the initial pairing stage ($\sim 200\,{\rm Myrs}$ since the first pericentric passage), the stalled systems generally experience less $\tau_{\rm grav}$, but not significantly.
We also compare the gravitational and dynamical friction torque from \texttt{ASTRID} (before the MBH merger) with the torque from the resimulation, to evaluate if we miss any influence of the large-scale structures on the MBH dynamics.
In general, we find that the $\tau_{\rm grav, resim}$ matches well with the $\tau_{\rm grav, ASTRID}$ values.
Because the dynamical mass of BH particles in \texttt{ASTRID} is boosted by $\sim 100$, its magnitude is closer to the gravitational torque and may have a larger impact on the dynamics.
This can also possibly lead to the early merger of the five systems that stalls in the resimulation.

Finally, Figure \ref{fig:energy} shows the change in the total energy of the secondary MBH after the first pericentric passage between the pairs, compared with the energy loss due to dynamical friction (in the resimulation).
We find a striking contrast between the merged and stalled system at the very early stage of the merger: the merged MBHs lost most of their energy within the first $\sim 100\,{\rm Myrs}$.
The energy loss due to dynamical friction then begins to take effect after $\sim 200\,{\rm Myrs}$ to further drive the merger.
In contrast, the stalled MBHs gained energy in this beginning phase from the gravitational torque. 
As a result, the MBHs never made their way into the central region where dynamical friction acts effectively.
On the right, we show a typical example of both the merged and stalled scenarios.
Consistent with the picture in Figure \ref{fig:tmerge_2d}, the MBHs in major mergers experience energy loss at the potential center of their hosts, whereas those in minor mergers with head-on collisions gain energy from the tidal disruption of their host galaxies.
In future development of subgrid merger models in cosmological simulations, it will be useful to measure the energy change rates of MBHs as an indication of the merging timescales and the likelihood of stalling.

\subsection{Effect of new infalling galaxies and MBHs}

The results shown for the seed MBH merging timescale so far exclude the effect of a third galaxy and MBH on the evolution of the original MBH pair.
Recall that in the system selection in Section \ref{sec:astrid_pop}, 6 systems (8-13) will start to have new infalling MBHs at $z\sim 6$, and two systems (14 and 15) are already in the multiple-MBH environment at $z\sim 9$.
In this section, we study the MBH pairing and orbits with considerations of multiple MBHs.

\begin{figure*}
    \centering
    \includegraphics[width=0.98\textwidth]{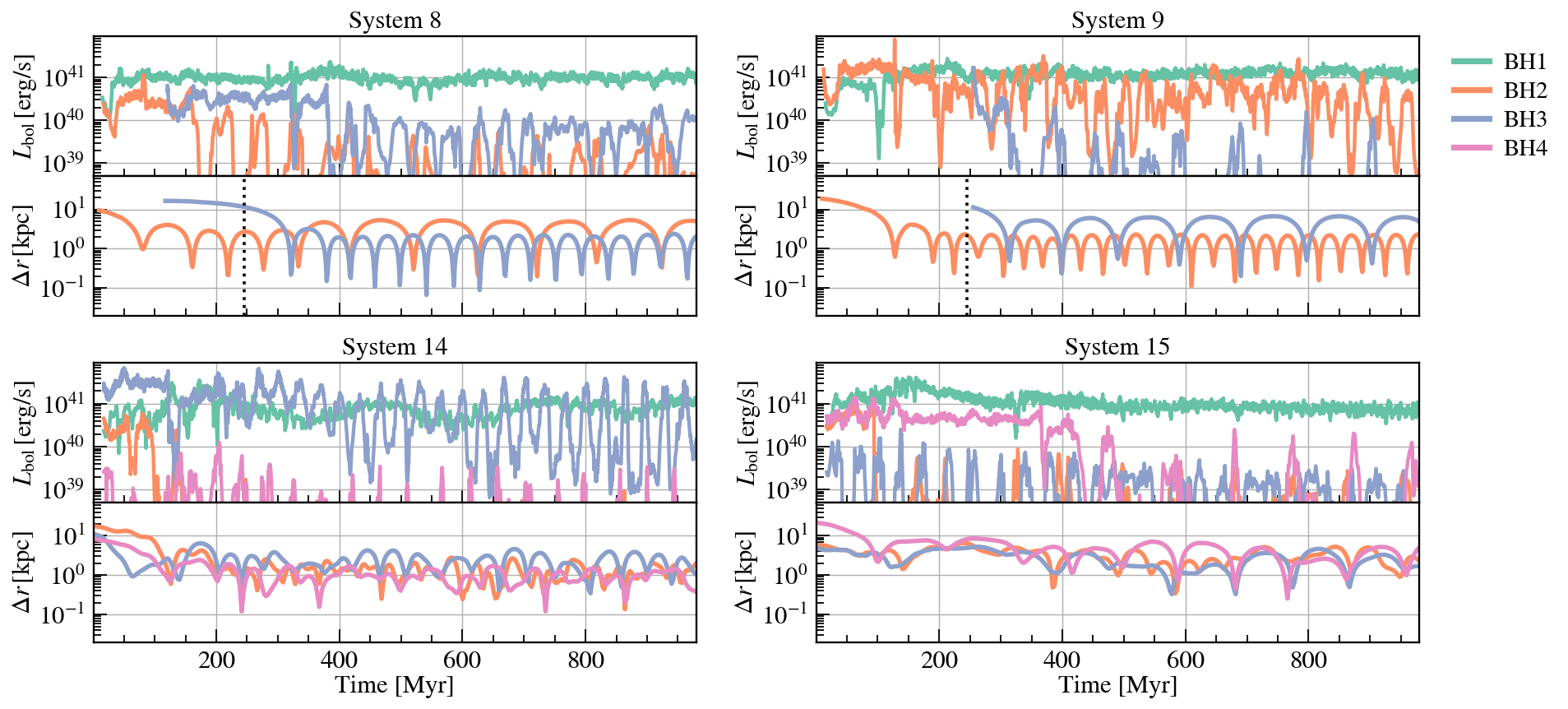}
    \caption{The bolometric luminosities and the orbits of all MBHs in four systems undergoing close interactions between multiple MBHs and galaxies. In all four cases, interactions between $>2$ galaxies and infalling MBHs lead to wandering MBHs on ${\rm kpc}$ scales.}
    \label{fig:orbits_inj}
\end{figure*}

Figure \ref{fig:orbits_inj} shows the bolometric luminosities and the orbits of all MBHs in four systems undergoing close interactions between multiple MBHs and galaxies.
In all four cases, we find stalling of all MBHs on kpc scales.
In particular, for the systems with new infalls (\texttt{system8} and \texttt{system9}), the third MBH/galaxy does not accelerate the sinking of the initial secondary.
In \texttt{system8} the secondary orbit widens with the infall of a new galaxy.
For the systems with simultaneous merger between several MBH-hosting galaxies, the MBHs exhibits more chaotic orbits except for the primary MBH.
These multiple MBH mergers lead to many wandering MBH seeds that do not grow efficiently in the remnant galaxy.
This picture is also consistent with the earlier findings that the seed dynamics are governed by large-scale torques.
The galaxy structure is often more complicated with changing potentials for the multiple merger case and can lead to energy increases of seed MBHs.

\subsection{Effect of nuclear star clusters}
\label{sec:taichi}

Recent works suggest that if MBHs are embedded in extended stellar systems such as NSCs, the sinking and formation of MBH binaries can be enhanced. This enhancement arises from the additional mass, which aids in dynamical friction, and the tidal effects exerted by the cluster \citep[e.g.,][]{Ogiya2020,Mukherjee2023MNRAS.518.4801M}. Our objective is to understand how resolving these clusters using N-body methods influences the outcomes obtained from the resimulations and compare the inspiral time obtained from the subgrid dynamical friction prescription used in {\tt\string ASTRID} resimulations.

We use the Fast Multipole Method (FMM) based N-body code {\tt\string Taichi} \citep[][]{Qirong2021NewA...8501481Z,Mukherjee2021ApJ...916....9M,Mukherjee2023MNRAS.518.4801M} to perform $N$-body simulations of the resimulated {\tt\string ASTRID} systems. {\tt\string Taichi} has explicit error control with time-symmetrized adaptive timesteps that allow the code to produce accurate results, even at mpc scales, and consistent with those obtained from direct summation-based $N$-body codes. {\tt\string Taichi} is highly efficient at simulating large-$N$ systems owing to the $\mathcal{O}(N)$ force calculations rather than $\mathcal{O}(N^2)$ that is typical of direct summation based $N$-body codes. 

We perform preliminary investigations of two systems - {\tt\string system5} and {\tt\string system8}. These systems are chosen since they lie at the density and energy criterion boundary separating merged systems from stalled ones. We are motivated to understand if embedding the MBHs in these systems in stellar clusters allows them to sink to sub-pc scales where the binary enters the hard-binary limit. 

The particle data is obtained at $t=968.2$ Myr for {\tt\string system5} and $t=500$ Myr for {\tt\string system8}. We take the particle data from the resimulations and perform radial cuts of 3 kpc and 5 kpc respectively from the centers of potentials of both systems. This was done to reduce the computational expenses. Cropping the systems results in $N=1.3\times10^6$ particles being retained from {\tt\string system5} and $N=2.4\times10^6$ particles being retained from {\tt\string system8}. We ensured that the cropping did not affect the overall dynamics of the MBHs and simulations were performed with non-cropped and cropped systems to verify consistency.

We infer the total mass present in clusters by extrapolating the stellar density profile obtained from the resimulations beyond 100 pc.
Since the mass is sensitive to the profile used for extrapolation, we use three different slopes to generate three different models for each system: a shallow cusp with $\rho(r) \propto r^{-1}$, a slightly steeper cusp with $\rho(r) \propto r^{-1.5}$, and a steep cusp with $\rho(r) \propto r^{-2}$. The mass of each cluster, $M_c$, is then calculated by subtracting the mass present within the inner 100 pc and dividing it by two.

To ensure that the masses of the clusters are physically realistic, we compare the initial cluster mass in each of the three models to the initial stellar mass present within 100 pc of the MBHs in each of the galaxies before they get disrupted. For {\tt\string system5}, we find that the lowest mass cluster is about $2\times$ the mass contained around the primary MBH. For the highest mass cluster, we find that the cluster mass is about $5\times$ that contained around the primary MBH initially. Similar values are obtained in the case of {\tt\string system8}. The lower initial mass inferred from the galaxies is caused due to the suppression in the density profile within 100 pc of the MBHs owing to softening. In general, we would expect cusps to form around the MBHs leading to a larger stellar mass which would be more consistent with the masses of the clusters that we used in this study. Additionally, we note that the total cluster mass to stellar mass in the galaxy ranges from 1-3\%, which is quite consistent with NSC to bulge stellar masses of some known nucleated dwarf galaxies \citep[e.g.,][]{Khan2021}.

Since {\tt\string Taichi} cannot handle gas effects, the gas particles are treated as stellar particles. We do not expect this to affect our overall results since gas is subdominant in the region of interest. $N$-body realizations of the stellar clusters are generated using the galactic modeling toolkit {\tt\string Agama} \citep{Vasilev2019MNRAS.482.1525V} by taking into account the potentials of the cluster, the MBH, and the galaxy. We use a Dehnen density profile \citep{Dehnen1993MNRAS.265..250D} to model the cluster with a shallow inner cusp of $\gamma=0.5$ and scale radius $a=1.4$ pc. All of the generated clusters have a half-mass radius of about $4.3$ pc. The cluster particles are assigned masses of $10^3 M_{\odot}$ each. Ideally, even smaller cluster particles are desirable to model the tidal effects accurately. While that is beyond the scope of this work, future work will include clusters that have a mass resolution of $10-100 M_{\odot}$. We summarize the initial conditions for our $N$-body simulations in Table \ref{tab:nsc_ic}.  

\begin{table}
\centering
\resizebox{0.45\textwidth}{!}{%
\begin{tabular}{lll}
\hline
System  & Cluster model  & $M_c [M_{\odot}]$   \\
\hline
{\tt\string system5} & \begin{tabular}[c]{@{}l@{}}Low mass\\ Intermediate mass\\ High mass\end{tabular} & \begin{tabular}[c]{@{}l@{}}$2.1\times10^6$\\ $3.1\times10^6$\\ $6.3\times10^6$\end{tabular} \\
\hline
{\tt\string system8} & \begin{tabular}[c]{@{}l@{}}Low mass\\ Intermediate mass\\ High mass\end{tabular} & \begin{tabular}[c]{@{}l@{}}$2.5\times10^6$\\ $3.7\times10^6$\\ $7.5\times10^6$  \end{tabular} \\ \hline
\end{tabular}%
}
\caption{A summary of the different stellar cluster models used in the $N$-body simulations and the masses of each individual cluster.}
\label{tab:nsc_ic}
\end{table}

\begin{figure}
    \begin{center}
    \includegraphics[width=0.45\textwidth]{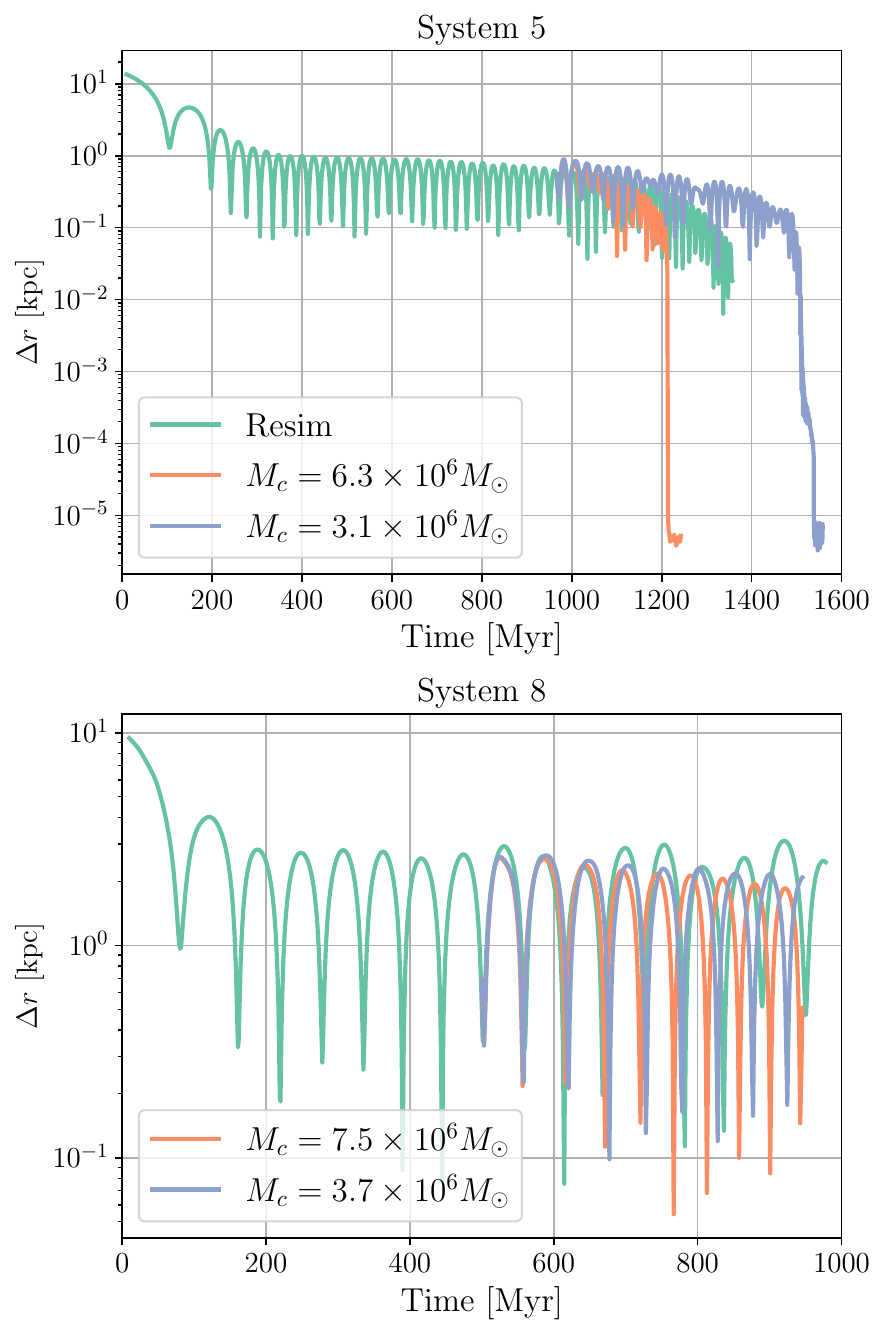}
    \caption{Relative separation between the MBHs $\Delta r$ as a function of the time for resimulations and models with MBHs embedded in NSCs for {\tt\string system5} (\textit{top}) and {\tt\string system8} (\textit{bottom}). In the high mass cluster model (\textit{orange}) in {\tt\string system5}, the MBHs can sink efficiently and form a hard binary by $1.2$ Gyr. The sinking time is almost twice as long for the intermediate mass cluster model (\textit{blue}) owing to the lower mass in the cluster. The DF prescription in the resimulations (\textit{green}) predicts an inspiral time somewhere within the two models.  In {\tt\string system8}, despite the added mass due to the clusters, the MBHs are unable to sink and form a hard binary. Even in the large mass cluster model, the separation between the MBHs reduces very slowly. This is quite consistent with the evolution in the resimulations.} 
    \label{fig:Taichi_inspiral_compare}
     \end{center}
\end{figure}

The systems are evolved for $\sim 500$ Myr beyond the initial time or until the formation of a hard binary. Plummer softening is used while calculating the forces. The softening used for the cluster particles is $0.01$ pc, while that for the stellar and gas particles is $25$ pc. When the separation of the MBHs decreases below 100 pc, we decrease the softening of the stellar and gas particles to 1 pc. Dark matter particles are assigned a softening length of $50$ pc. The interactions between the MBHs are never softened. The softening lengths were varied to understand the effects on the sinking time and no major differences were noticed. In scenarios that result in the sinking of the MBHs to sub-pc length scales, convergence is ensured by running the simulations again after splitting the particles such that the overall mass resolution of the non-cluster particles is $2\times10^3 M_{\odot}$. Particle splitting is performed using the same procedure as used in some previous studies \citep[e.g.,][]{Khan2012ApJ...756...30K}. In the split-particle cases, the softening of the DM particles is reduced to $25$ pc. For {\tt\string system5}, particle splitting results in a total of $N \approx 3\times 10^6$ particles. While the mass resolution used in this work is somewhat insufficient to resolve the three-body hardening phase accurately, we want to note that the main objective of this preliminary study is to compare the sinking timescales between the ASTRID resimulations and the $N$-body simulations. A more detailed analysis is in preparation which includes additional prescriptions for relativistic effects, MBH spin, and GW recoil.

We use the fourth-order hierarchical Hamiltonian splitting integrator {\tt\string HHS-FSI} \citep{Rantala2021MNRAS.502.5546R}. We set a force error tolerance parameter of $\epsilon=2\times10^{-5}$, multipole parameter of $p=12$, and timestep parameter of $\eta=0.3$. This results in an overall relative energy error of $\sim 10^{-5}$ at the end of the simulations. For more information on the parameters, we refer the interested reader to \cite{Mukherjee2021ApJ...916....9M}. The simulations are run using 32-48 threads on a single {\tt\string AMD Epyc 7742} machine.

\begin{figure}
    \begin{center}
    \includegraphics[width=0.45\textwidth]{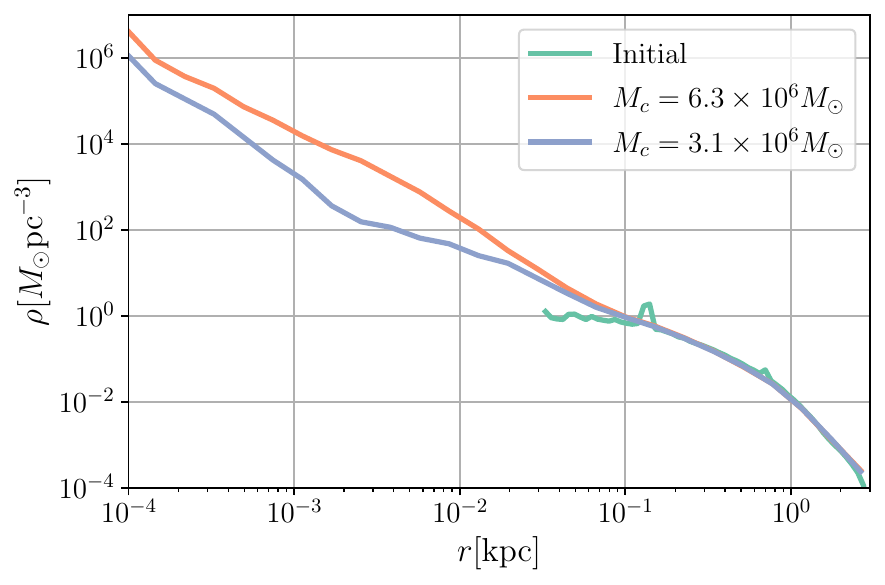}
    \caption{The stellar density profile upon the formation of a hard binary in the intermediate mass (\textit{blue}) and high mass cluster models (\textit{orange}) in {\tt\string system5}. The bumps in the initial stellar density profile (green) represent the positions of the NSCs initially. Consistent with our initial conditions, the density profile after the clusters have sunk form a $\rho(r) \propto r^{-1.5}$ profile in the inner 100 pc in the intermediate mass model and a $\rho(r) \propto r^{-2}$ profile in the high mass model. The central density at $10^{-4}$ kpc is quite consistent with stellar density values of known nucleated dwarf galaxies such as M32 or NGC 5102. } 
    \label{fig:Taichi_stellar_profile}
     \end{center}
\end{figure}

Examining Figure \ref{fig:Taichi_inspiral_compare} where we plot the relative separation of the MBHs $\Delta r$ as a function of time for our different $N$-body models, we find interesting results. For {\tt\string system5}, our high mass cluster helps the MBHs efficiently sink to the potential minimum of the galaxy and form a hard binary within 1.2 Gyr. The initial periastron separation between the MBHs is $\approx 800$ pc. The separation drops primarily due to DF on the extended system until the MBHs have a $\sim 50$ pc separation after which tidal forces from the clusters helps the MBHs sink rapidly to sub-pc separations, consistent with previous studies \citep{Ogiya2020,Mukherjee2023MNRAS.518.4801M}. The semi-major axis of the formed binary is around $5\times10^{-3} \rm{pc}$ whereas its eccentricity is $\sim 0.3-0.4$. We caution the reader, however, that a better estimation of eccentricity would require finer resolution in the last few Myrs before sinking, a work in progress. In the intermediate mass cluster model, the inspiral takes twice as long owing to the lower mass of the cluster, and some mass loss from the cluster due to tidal stripping. Similar to the high mass model, the separation between the MBHs rapidly drops when the separation reduces to $\leq 50$ pc, and a hard binary is formed quite efficiently. Although not presented here, we find that the low mass cluster model is not able to sink the MBHs to sub-pc scales within $\sim 500$ Myr of evolution but a decrease in the separation is noticed. We notice that the inspiral time predicted by the {\tt\string ASTRID} resimulation model is approximately in between our intermediate and large cluster models.

The cluster models in {\tt\string system8} do not show the same signs of rapid inspiral as those observed in {\tt\string system5}. Even in the high mass cluster model {\tt\string system8}, the MBHs are only able to reduce their periastron separation from 2.6 kpc to 2 kpc within 500 Myr. The decrease is even smaller in the case of the intermediate mass cluster model. Since {\tt\string system5} and {\tt\string system8} share similar initial MBH orbits and stellar density profiles, it suggests that NSCs become an effective method of sinking only when the periastron separation between the MBHs is $\lesssim 1$ kpc. This underscores the importance of taking into account the global effects and the necessity of {\tt\string ASTRID}-like simulations where the initial orbits of the MBHs are modeled accurately. The first few kpc scale orbits contain very useful information on the subsequent orbital evolution and the fate of the binary.

We also examine the overall stellar profile of the galaxy once a hard binary has formed in Figure \ref{fig:Taichi_stellar_profile}. Consistent with the initial conditions used, once the clusters have merged, we find that a $\rho(r) \propto r^{-1.5}$ profile forms in the inner 100 pc in the intermediate mass cluster model while a $\rho(r) \propto r^{-2}$ profile forms in the high mass cluster model. Since the softening of the cluster particles is quite small and they dominate the mass at $r \lesssim 0.1$ kpc, the density profile is accurate to $\approx 10^{-4}$ kpc. The stellar profile is quite consistent with those from known nucleated dwarf galaxies such as M32, NGC 5012, and NGC 5206 \citep{Khan2021}, especially in the inner-pc. The stellar profiles are obtained at the beginning of the hard-binary stage and no core scouring has taken place. With time, due to core-scouring, we expect the density within the influence radius of the binary ($\sim 1 \mathrm{pc}$) to become shallower.

%% file: Sec6_Conclusion.tex
 \section{Discussion}
 The dynamics of MBHs in the dynamical friction regime (from a few kpc to $<10\,{\rm pc}$ scales) have not been understood in great detail, partly due to the intrinsic stochasticity \citep[e.g.][]{Nasim2020, Rawlings2023} and the wide range of physics processed involved \citep[e.g.][]{Tamburello2017, Dosopoulou2017, Park2017, Banik2021}.
 Recently, several works have been focusing on bridging this gap in the dynamical range by following the dynamics of MBH pairs from galaxy mergers to binary hardening, and some even to binary coalescence.
 Here we briefly discuss our work in the context of these emerging literatures.

 Probably most relevant to our work are the recent studies by \cite{Koehn2023} and \cite{Partmann2023}, both of which take into account realistic consecutive galaxy mergers with $>2$ MBHs.
 The former uses a similar resimulation approach to resimulate triple SMBHs in the \texttt{Romulus} simulation, with a focus on the massive galaxies.
 The latter considers the infall of several seed MBHs in satellite halos into the main halo, and finds that a seed mass of $\sim 10^5\,M_\odot$ is needed for MBHs to merge in these low-mass galaxies.
 Compared to these works, our study uniquely considers the effect of gas physics in high-redshift galaxies, and shows that the growth in stellar density due to star formation allows seed MBHs to enter into the hardening phase within $\sim 1\,{\rm Gyr}$.
 However, we currently lack the self-consistent treatments of triple MBH interactions and gravitational recoil, which have been shown to lead to a high fraction of MBH ejections according to these works.
 Despite the differences in the subgrid models, the three works reach agreements on the production of numerous wandering MBHs due to various mechanisms in multiple-MBH systems.
 We also explicitly show that such systems are numerous as a result of the high-redshift early galaxy assemblies, and so it is important to understand such systems in greater detail and implications for early MBH growths.

 Other works on the seed MBH dynamics include \cite{Tamfal2018}, \cite{Pfister2019}, \cite{Ma2021}.
 In particular, both \cite{Pfister2019} and \cite{Ma2021} considered the effect of clumpy gas on the sinking of MBH seeds.
 The key finding is that seed-mass MBHs cannot sink efficiently in clumpy high-redshift galaxies, or even in idealized cases.
 By taking initial conditions from cosmologically merged systems instead of putting MBHs on more \textit{ad hoc} orbits, our resimulations naturally favor the initial orbital parameters and galaxy configurations (such as rotation angles) that are more likely to sink the MBHs efficiently.
 With initial conditions that favor efficient orbital decay, it is still likely that the MBHs can migrate into the dense central star-forming regions before they get significantly scattered.
 In future works, we will explore if more detailed ISM modeling significantly impacts our conclusion.

\section{Conclusion}

We present a suite of the MAGICS simulations, consisting of 15 idealized high-resolution galaxy merger simulations with initial conditions directly reproducing the configurations of galaxy mergers in the large-volume cosmological simulation \texttt{ASTRID}.
This suite encapsulates a wide range of realistic galaxy merger environments directly drawn from \texttt{ASTRID}, with both isolated dwarf galaxy mergers with an MBH pair and consecutive galaxy mergers with multiple MBHs.
The simulation suite is run with the full subgrid physics model of the \texttt{ASTRID} simulation using \texttt{MP-Gadget} to include realistic gas and star-formation physics, the feedback from the supernova, as well as the accretion onto MBHs and the AGN feedback.

We use these simulations to study the merger and sinking of high-redshift ($z\sim 6$) mergers between MBH seeds ($5\times 10^4\, M_\odot<M_{\rm BH}<10^6\,M_\odot$), in early gas-rich, star-forming galaxies with a typical gas fraction $>1$, galaxy mass between $\sim 10^8\,M_\odot$ and $\sim 3 \times 10^9\,M_\odot$, and halo mass between $\sim 5\times 10^{10}\,M_\odot$ and $\sim 5 \times 10^{11}\,M_\odot$.
For selected systems with inefficient orbital decay, we continue the galaxy merger simulation during the evolution of the final $\sim {\rm kpc}$ scale with the FMM-based N-body code \texttt{Taichi} with MBHs embedded in NSCs.

We show that the resimulation method can create galaxy merger initial conditions and orbital configurations that resemble the original galaxy merger system in cosmological simulation.
Furthermore, the subsequent evolution of the cosmological merger is also paralleled in the resimulations regarding the MBH orbits, host galaxy density profiles, and host galaxy star formation rates.
We find a good match between the MBH orbits of the \texttt{ASTRID} mergers and the resimulation mergers above the \texttt{ASTRID} resolution limit (the first $\sim 3$ orbits).
This validates using cosmological simulations with dynamical friction modeling for setting the initial distribution of MBH orbital eccentricities and separations for idealized galaxy/MBH mergers simulations and analytical models.

In 8 out of 15 resimulated \texttt{ASTRID} mergers, the MBH pairs can sink efficiently to separations below $20\,{\rm pc}$ in $1.5\,{\rm Gyrs}$ and before other galaxies start to interfere with the binary system.
\texttt{ASTRID} mergers with high initial eccentricity ($e_{\rm init}>0.5$), high density at kpc scales ($\rho_{\rm star}>0.05\,M_\odot/pc^3$), and low halo mass ratio ($q_{\rm halo} < 0.3$) will sink efficiently to $\sim 20\,{\rm pc}$ in the resimulation.
The MBH will stall at $0.1-1\,{\rm kpc}$ orbits if any of these conditions are not satisfied.
Moreover, the central stellar density can grow by a factor of $2\sim 3$ between the cosmological merger and the end of the dynamical friction regime and over a factor of $\sim 10$ from before the galaxy merger to the binary hardening.
% These systems are isolated galaxy mergers characterized by high star-formation rates, stellar densities, and high eccentricity.
We find that all merger remnant galaxies have a central stellar density of $\rho_{\rm star, 50pc} > 1\,M_\odot/pc^3$ and orbital eccentricity of $\sim 0.8$ when the MBHs begin entering into the binary hardening regime. 
Our predicted eccentricity is in broad agreement with the results in \cite{Gualandris2022}.

By directly linking the resimulation MBH mergers (or non-mergers) with the cosmological system, we find that galaxy and orbital properties at the \texttt{ASTRID} MBH merger time and resolution are already good indicators of whether the MBH pair can sink efficiently or not.
Specifically, the energy loss of the seed MBH that leads to fast sinking is dominated by gravitational torque during the first $\sim 200\,{\rm Myrs}$ of galaxy merger, and dynamical friction plays a subdominant role.
The seed MBHs in merged systems lose energy at a rate of $\sim 10-100\,{\rm km\,s^{-2}\,Myr^{-1}}$ during the galaxy merger, whereas the stalled seeds gain energy in this phase.
As a result of the initial energy loss driven by large-scale torques, some seeds experience efficient orbital decay during the first few orbits of MBH paring, with apocentric orbital sizes below $1\,{\rm kpc}$ in $\sim 200\,{\rm Myrs}$ after the first pericentric passage between the MBHs.

Consecutive mergers between multiple galaxies and MBHs are common ($\sim 50\%$) among high-redshift seed MBH mergers, and thus need to be taken into account when modeling the merging between MBH seeds.
We find that the consecutive merger scenario generally hinders the sinking of MBH seeds.
The 4 resimulated systems with multiple galaxy mergers involving $>2$ MBHs all lead to the stalling of several MBHs at $\sim 1\,{\rm kpc}$ from the remnant galaxy center, with only one MBH sinking and accreting efficiently at the galaxy center.
In particular, we find that the orbit of the initial MBH pairs widens with the infall of new galaxies.

Finally, by resimulating the sub-kpc evolution of MBH pairs for two systems with the secondary MBHs stalling on $\sim 1\,{\rm kpc}$ scales with MBH embedded in NSCs, we find that a cluster mass of $\sim 3\times 10^6\,M_\odot$ facilitate the sinking of the secondary and allows for a rapid formation of a hard binary in the case where the orbital size is already below $1\,{\rm kpc}$. 
By applying the criterion of rapid MBH sinking derived in this work (high-density and high-eccentricity mergers in isolation) to the properties of \texttt{ASTRID} MBH mergers shown in \cite{Chen2022}, we find that about $\sim 10-20\%$ MBH seeds that pair at $z\sim 9$ will enter into the binary hardening phase before $z\sim 3$.
Since binary hardening is also relatively efficient for dense and high-eccentricity mergers, we expect that these pairs will be detected by LISA around $z\sim 3$.

%% file: AppA_convergence.tex
\section{Initial condition parameters}
\label{app:ic}
In Table \ref{tab:ic}, we show more detailed decomposition of each galaxy hosting a seed MBH in our initial conditions, as recovered from the galaxy properties in \texttt{ASTRID}.
Most galaxies are bulge-dominated, and with halo gas component dominates over the disk gas component.
Nonetheless, the disk gas is still more massive than the stellar component in the initial conditions.
The large fraction of disk gas will lead to rapid star formation when galaxies merge.
%%%%%%%%%%%%%%%%%%%%%%%%%%%%%%%%%
\begin{table*}
\centering
\caption{The masses and sizes of each component in the initial conditions of merging galaxies. In the case of multiple galaxies or MBH mergers, we record the information of two most massive halos hosting the MBHs.}
\label{tab:ic}
\begin{tabular}{lcccccccccc}
\hline
Name & $M_{\rm bulge\,1,2}$  & $M_{\rm disk\,1, 2}$ & $M_{\rm disk\,gas\,1, 2}$ & $M_{\rm halo\,gas\,1, 2}$ &  ${\rm disk\,scale\,1, 2}$ & ${\rm bulge\,scale\,1, 2}$  \\
& [${\rm M}_\odot$]  & [${\rm M}_\odot$]  & [${\rm M}_\odot$]  & [${\rm M}_\odot$]  & [${\rm kpc}$] & [${\rm kpc}$]  \\
\hline
\texttt{system1} & 2.6e7, 5.1e6 & 1.4e7, 2.2e6 & 8.6e8, 3.9e8 & 2.2e9,  1.6e9 & 0.85, 0.47  & 0.23, 0.22 \\
\texttt{system2} & 1.3e8,  4.2e7 & 7.2e7,  1.9e7 & 2.4e8, 2.4e8 & 4.5e9, 4.5e9 & 0.58, 0.63 & 0.30, 0.58\\
\texttt{system3} & 4.8e6, 3.8e6 & 4.4e6, 6.4e6 & 4.9e8, 2.6e8 & 1.1e9, 1.0e9 & 0.13, 0.15 & 0.26, 0.41\\
\texttt{system4} & 2.7e7, 1.0e7& 9.6e6, 5.4e6 & 1.3e9, 5.6e8 & 3.8e9, 7.6e8& 0.86, 0.17& 0.44, 0.34\\
\texttt{system5} & 9.3e6,  5.5e6 & 4.3e6,  5.2e6 & 1.7e8,  5.1e7 & 3.2e9, 9.8e8 & 0.39, 0.35 & 0.32, 0.34 \\
\texttt{system6} & 3.4e6,  2.0e6& 4.2e6, 2.4e6 & 4.1e8, 3.5e8& 7.4e8, 5.0e8 & 0.24,  0.22& 0.26, 0.24\\
\texttt{system7} & 1.2e7, 3.9e6 & 9.0e6, 2.9e6& 5.8e8, 1.9e8& 1.6e9, 9.9e8& 0.23, 0.16 & 0.31, 0.28 \\
\hline
\texttt{system8} & 1.5e7, 4.7e6& 8.8e6,  3.4e6& 1.5e8, 2.8e7 & 2.9e9, 5.4e8 & 0.17, 0.11 & 0.22, 0.2\\
\texttt{system9} & 1.0e6, 5.0e6 & 1.8e6, 2.1e6 & 8.0e7, 7.1e7 & 1.5e9, 1.3e9& 0.16, 0.38& 0.33, 0.26\\
\texttt{system10} & 1.6e7,  1.7e7& 1.3e7, 1.7e7 & 1.7e9, 2.1e9& 3.9e9, 2.1e9 & 0.94, 0.56& 0.36, 0.46\\
\texttt{system11} & 1.6e8, 2.1e7 & 5.7e7, 1.4e7 & 8.0e8, 2.3e8& 1.5e10, 4.3e9 & 2.9, 1.1& 0.30, 0.48 \\
\texttt{system12} & 1.1e7, 8.5e6 & 8.3e6, 7.1e6& 1.3e8, 7.4e7 & 2.8e9, 1.4e9 & 0.33, 0.30& 0.40, 0.27 \\
\texttt{system13} & 8.3e6, 6.3e6 & 4.1e6, 2.8e6& 5.2e8, 2.3e8 & 1.4e9,  3.5e8& 0.39, 0.13 & 0.26, 0.23\\
\hline
\texttt{system14} & 9.3e7, 1.1e7 & 4.7e7, 6.7e6 & 3.2e9, 3.6e8& 6.8e9, 1.9e8& 1.3,  0.15& 0.34, 0.30 \\
\texttt{system15} & 1.2e8, 9.3e6 & 8.4e7, 5.5e6 & 6.4e8, 4.8e7& 1.2e10, 9.1e8& 0.88, 0.32& 0.68,  0.33\\
\hline
\end{tabular}
\end{table*}